# Welfare Reform: Consequences for the Children

Marianne Simonsen
Department of Economics
and Business Economics
Aarhus University

Lars Skipper
Department of Economics
and Business Economics
Aarhus University

Jeffrey A. Smith
Department of Economics
University of Wisconsin-
Madison

December 17, 2025

*Abstract*

This paper uses register-based data to analyze the consequences of a recent major Danish welfare reform for children's academic performance and well-being. In addition to work requirements, the reform brought about considerable reductions in welfare transfers. We implement a comparative event study that contrasts outcomes for individuals on welfare at the time of reform announcement before and after the implementation of the reform with the parallel development in outcomes for an uncontaminated comparison group, namely those on welfare exactly one year prior. Our analysis documents that mothers' propensity to receive welfare decreased somewhat as a consequence of the reform, just as we observe a small increase in hours worked. At the same time, we do not detect negative effects on children's short-run academic performance. We do find small negative effects on children's self-reported school well-being and document substantial upticks in reports to child protective services for children exposed to the reform.

---

Acknowledgements: We appreciate comments from Jeff Hicks, Helena Skyt Nielsen, Marianne Page, and Emilia Simeonova. We also received valuable inputs from participants in the 5th Family and Education Workshop, the 2020 Copenhagen Education Network Workshop, the 2021 IRP Summer Research Workshop, the 2021 IIPF Annual Congress, the 2022 CESifo Area Conference on the Economics of Education, and the 2023 Canadian Economic Association Meeting, as well as seminar participants at SOFI, IFN, University of Erlangen-Nuremberg, Stavanger University, and Aarhus University. We gratefully acknowledge funding from the Independent Research Fund Denmark. The usual disclaimer applies.

## 1. *Introduction*

Over the last few decades, many governments – including the Anglo-Saxon and the Nordic welfare states – have reformed welfare systems with the purpose of promoting work (e.g., Grogger and Karoly, 2005; Mogstad and Prozanto, 2012), for example by introducing work requirements backed up by sanctions or by increasing financial work incentives. Critics worry that while welfare reforms do lead some to work (or to work more), others may simply experience income loss. Moreover, the literature has long recognized (e.g., Holmes and Rahe, 1967; Milligan and Stabile, 2011) that family financial shocks stress both parents and children. As such, we might expect welfare reform to have broader implications within households, which makes estimates of effects on partner and (especially) children's outcomes an important part of a complete accounting of the costs and benefits of reform.

This paper uses population-wide register data on parents' welfare participation coupled with a range of parental, partner, and child outcomes across several domains to study a welfare reform that introduced an upper limit on welfare benefits as well as work requirements. Our identifying variation comes from a recent major Danish reform passed into law in March 2016 and implemented in October 2016. The reform imposed a substantial policy change by introducing work requirements that, if not fulfilled, could lead welfare recipients to lose all or part of their benefits. At the same time, absent any behavioral responses, the upper limit on welfare benefits reduced disposable income for welfare recipients by five to 20 percent depending on family type (The Danish Ministry of Employment, 2015a). The reform affected a large population: roughly 170,000 Danes received welfare benefits at the time of the announcement and more than 30,000 children subsequently experienced a reduction in their families' benefits (Statistics Denmark, 2017).

Relative to the existing literature in the area, we utilize high-frequency panel data on a rich palette of outcomes in a comparative event study (i.e., difference-in-differences with treatment effects varying with time relative to the policy change) that exploits the exact timing of the reform. Specifically, to avoid issues with anticipation we start by selecting the group of mothers on welfare in the month in which the reform was passed into law. Our strategy then compares the change in outcomes for individuals in the reform group around the policy change with the corresponding development in outcomes for a comparison group of individuals on welfare exactly one year prior. This strategy allows us to speak to the effects of the reform for children with mothers on welfare at the time of the reform. Our approach has the advantage that it balances out within-calendar-year dynamics in welfare participation. Zeroing in on variation generated by the timing of the reform also minimizes the role of other, concurrent factors. These advantages in turn lead to the disadvantage that we can look at impacts only in the first year following the reform, after which point our comparison group receives the reform treatment as well. Our estimates may understate long-run changes if adjustments take time, or may overstate long-run changes to the extent that they capture short-run adjustment costs (e.g., stress, disagreements about the allocation of the burden within the household, and/or residential mobility) that fade with time.

To illustrate the immediate workings of the reform, Figure 1 shows monthly government transfers before and after the reform for the population of mothers on welfare benefits in March 2016. The figure then compares these to the transfers paid out exactly one year earlier but to the population of mothers on welfare in March 2015 – the comparison group in our empirical strategy. The benefit levels of the two populations track each other closely long before the reform was passed into law and continuing until the reform was implemented in October 2016 for the reform population. At that point, we observe a sudden decrease in the value of government transfers of about 12 percentage points.

The overall response illustrated in Figure 1 combines a small extensive margin response and a larger intensive margin response. In particular, we show that mothers' propensities to receive welfare decreased only modestly as a consequence of the reform, just as we observe only a small increase in hours worked. As a result of the reform, average monthly household discretionary income[1] fell by just above € 300.[2] We do not detect effects on short-run child academic performance, but do find small negative effects on children's self-reported school well-being, as measured by individual-level nationally administered surveys. Moreover, we document substantial upticks in reports to child protective services for children exposed to the reform. These reports typically concerned child externalizing behaviors, insufficient care by parents, or high levels of family conflict. Analyses of supplementary outcomes related to school absence rates and the prevalence of injuries support our main conclusions.

[FIGURE 1 AROUND HERE]

Our strategy relies, among other things, on the absence of concurrent changes. Having first confirmed that any effects on the inflow into welfare were miniscule, we therefore investigate trends in outcomes in the non-welfare population. We find that our main conclusions with regards to mothers' outcomes still stand when we implement a triple comparative event study that accounts for overall trends in outcomes in the non-welfare population too. One year after the passage of the reform, we estimate an effect on mothers' hours worked of 3.9 per month with our main strategy and 3.5 when we rely on the triple comparative event study. The corresponding estimates for receipt of welfare are -8.2% and -7.9%. All estimates are statistically significant at a 5%-level.[3] For children, the estimated effect on Danish reading turns statistically significant at the 5%-level and increases from 0.015 standard deviations to 0.048 standard deviations in the triple comparative event study. In contrast, the effect on social well-being becomes essentially zero. We reach similar conclusions for math and reports to child protective services across specifications; in both cases, a small and statistically insignificant effect on math and meaningful effects on reports to child protective service that differ only at the third digit after the decimal point. Overall, our estimates paint the welfare reform as harmful to children on average, at least in the short run, but far from a disaster.

[1] The sum of mother's and any partner's earnings and transfers minus taxes and rent.
[2] The Danish krone is pegged to the euro with a DKK/EUR exchange rate of 7.46. As of August 16, 2025, the USD/EUR exchange rate is 1.17.

We structure the remainder of the paper as follows: Section 2 describes prior knowledge on the link between welfare reform and child outcomes, while Section 3 concerns itself with the institutional setting as well as the content of the welfare reform. Section 4 explains our empirial strategy and Section 5 presents the data, while Section 6 shows the results and Section 7 concludes.

## 2. *The link between welfare reform and child outcomes*

The direction of the effects of welfare reform that aims at promoting work, often through reductions in cash benefit levels and via work requirements, on child outcomes is ex ante unclear. To the extent that the tightening of welfare policies results in lower family income and more stress in the family, we expect that children will be harmed by such reforms. A number of papers (e.g., Dahl and Lochner, 2012; Aizer et al., 2016; Akee et al., 2018, and Akee et al., 2024) document that income matters for children's emotional and behavioral health as well as later life human capital outcomes, perhaps because higher income results in better home environments (e.g., Cesur et al. 2025). More broadly, Mayer (1997) takes a skeptical early look at the household determinants of child outcomes, Almond and Currie (2010) review the literature showing that early stressors (both monetary and not) lead to worse child outcomes, and Page (2024) provides an admirable recent survey focused more narrowly on financial resources.[4]

[3] We follow the literature in pretending we have a sample even though our data comprise the population of interest. The notion of super-populations may comfort some readers in this regard. More traditional readers, who know a population when they see one, may ignore the standard errors.

Still, some parents might take up work because of the reforms, which holds the potential to benefit children, for example because working parents can serve as positive role models for their children, and because of the earned income (e.g., Heinrich, 2014). However, the empirical evidence on the causal link between parental (mainly maternal) employment and child outcomes remains varied and contentious, though researchers generally agree that the effect varies with child age and socio-economic background (e.g., Berger et al., 2005; Ruhm, 2004; Ruhm, 2008; Herbst, 2017).

Uncovering the effects of various welfare reforms on child outcomes has proven difficult because of data disconnects between children and the relevant adults; lack of consistent data on relevant child outcomes across age, time, and geographic space; small samples and issues with survey attrition; and the occurrence of simultaneous policy changes in related programs. Our paper relates directly to a smaller literature concerned with the link between the tightening of welfare and child human capital development. To the best of our knowledge, most other studies that look at child outcomes concern the 1990s US welfare reforms that introduced various combinations of time limits on welfare receipt, benefit changes, wage subsidies, work requirements, and childcare subsidies. The US welfare reform period also saw considerable expansions of the Earned Income Tax Credit, which targets a similar population and has an independent effect on adult labor supply (e.g. Eissa and Hoynes, 2004). Some of the state-level US welfare reforms (done under waivers prior to the national reform) did undergo experimental evaluation but given the policy focus on adult employment and difficulties in administrative data linkage, many evaluations did not consider child outcomes. The ones that did offer evidence of both positive and negative effects on child well-being; see Grogger and Karoly (2005) for an overview and Gueron and Ralston (2013) for a larger evaluation-oriented history of welfare reform. More broadly, the combination of state-level heterogeneity in program details and implementation with the multi-dimensional nature of US welfare reform largely derails comparisons of the magnitudes of treatment effect estimates.

Nonetheless, we highlight three studies look at effects of US welfare reform on child outcomes, along with two Canadian studies and a Norwegian one. We priviledge these six based on their topical similarity, their recency, and the quality of the evidence they provide. In the U.S. context, Miller and Zhang (2012) were the first to measure the impact of welfare reform on the educational attainment of children in low-income families using large, nationally representative samples. They use a version of difference-in-differences that compares children of low-income parents with children of higher income parents before and after the reforms. Perhaps surprisingly, they find that income gaps in school enrollment and drop-out rates narrowed by more than 20 percent as a consequence of the reforms. Reichman et al. (2020) exploit state and time variation in the US welfare reforms and document modest but meaningful reductions in mothers' engagement with their children, especially boys. Kalil et al. (2023) use a similar strategy and find reductions in maternal emotional support of 0.3-0.4 standard deviations.

[4] A small recent literature ably summarized by Smeeding (2024) looks at the effects of the temporary increase in the US Child Tax Credit (CTC) during the damn pandemic on child and adult outcomes. In our view, the transitory nature of the CTC treatment and the unusual context make this literature less directly relevant to our work.

Only a few studies employ data from outside the US. The paper closest to ours in terms of child data availability is Hicks et al. (2023). That paper exploits a particular feature of a 2002 welfare reform in the Canadian province of British Columbia that reduced the age of the youngest child at which a mother on welfare was required to search for work. Their empirical strategy essentially compares families where the youngest child was aged 4 to 6 (the treatment families) to families wherein the mother was required to search for work both before and after to the reform; namely those with a youngest child aged 8 to 11. They find no effects on grade 10 test scores, on high school graduation, or on criminal charges as a young adult. They do find that welfare receipt as a child increases the probability of welfare receipt in early adulthood. In addition to the different settings, our paper clearly distinguishes itself from Hicks et al. (2023) in the nature of the parameter of interest. Rather than considering the effects of welfare enrollment per se, we (mainly) consider the effect of a change in benefit levels for those remaining on welfare, in the context of a reform with little effect on employment or the extensive margin of welfare receipt.

Morris and Michalopoulos (2000) provide experimental results from the Canadian Self-Sufficiency Project (SSP), which offered a temporary but generous earnings supplement to lone mothers in two provinces conditional on full time work. Their analysis of outcomes 36 months after random assignment (RA) yields effects that vary with child age. Specifically, though the program increased take-up of child care, they find no effects on test scores, social behavior, emotional well-being, or health for children under three years old at RA. For children aged three to eight, the study shows some positive effects on math skills test performance. Results for children ages nine to 15 at RA show increases in minor delinquency and in tobacco, alcohol, and drug use, though the authors worry about their lower survey response rates among older children. Again, unlike in our paper, the extensive margin of social assistance partipation plays a major role in their story.[5]

Finally, Løken et al. (2018) investigate the effects of a 1998 Norwegian reform targeted at single mothers. The reform was implemented over a period of three years and had three main components: work requirements, a reduction in the maximum period of benefit receipt from nine to three years, and a slight increase in benefit levels. As ever, the portmanteau nature of the treatment complicates forming an informed prior about its effects as well as comparing its effects to those of other reforms. The study uses a difference-in-difference strategy that compares the outcomes of children of single mothers with those of married mothers. It finds a small but detectable drop in school grades at age 16 by 0.7% of a standard deviation per additional year of maternal exposure to the reform.

In sum, the existing evidence base regarding the effects of welfare benefits on child outcomes is small, based on a variety of mostly multi-dimensional policy changes, and mixed in its findings. It awaits a through survey and reconciliation. To this, we add estimates based on relatively clean (and relatively recent) variation in benefit levels; put differently, we examine a change in benefit levels with almost no extensive margin effects on welfare receipt or unemployment levels, and with no additional policy dimensions (e.g., child care subsidies or time limits or job search requirements) to complicate interpretation. In addition, we offer compelling causal identification, and provide evidence on effects on both usual suspect outcomes and on outcomes generally absent from the existing literature due to lack of available data.[6,7]

[5] A different vein of work concerns itself with intergenerational spillovers in welfare dependence. A recent paper by Hartley et al. (2022) shows that the US welfare reforms of the 1990s weakened the link between mothers' and daughters' welfare participation. Another recent paper by Dahl and Gielen (2021) finds that limiting access to disability insurance for parents in the Netherlands reduces their children's take-up of disability insurance and improves a range of other outcomes.

[6] Recent research has studied welfare reforms related to the population of refugees. Dustmann et al. (2024) study a Danish 2003 reform that reduced benefits to refugee immigrants (more than half of which arrived from Iraq) by around 50 percent for those granted residency after the reform date. They show that childrens' performance in language tests as well as length of education decreased as a consequence of the reform, just as teenagers' crime rates increased. Similarly, Andersen (2024) studies a Danish 2015 reform that gave refugees, who were primarily from Syria (83%), lower benefits if they were granted residency after September 2015. She shows that the reform increased children's absence from primary school. As described in Section 5, our population of interest excludes recent migrants, precisely because they experience the alternative transfer programs analyzed in these papers.

[7] Another set of recent papers looks at the long-term effects of exposure to safety net programs as a child, for example the introduction of the Mothers' Pension program (1911-1935) or the county-level roll-out of the Food Stamps program between 1961 and 1975. This set includes Aizer et al. (2016), Bastian and Michelmore (2018) and Bailey et al. (2020). This literature tends to find gains from access in terms of child human capital accumulation. The temporal external validity of these estimates to the present day remains an open question, as the relevant counterfactual has often changed substantially in the interim.

## 3. *The 2016 reform*

According to the Danish Law of Active Social Policy ("Lov om Aktiv Socialpolitik"), individuals qualify for welfare benefits in cases of job loss or prolonged illness if they cannot provide for themselves and their families through other income sources, such as unemployment insurance, or by depleting their assets.[8] The benefits include welfare payments ("kontanthjælp") and housing support. Welfare benefits increase discontinuously with age, with higher payments to individuals over 30, and vary by family size, with additional top-ups for single parents. Housing support primarily consists of general housing assistance ("boligsikring") a means-tested but universal benefit not limited to social housing, supplemented by special support ("særlig støtte").[9]

One issue is that financial incentives to find work are limited due to a high benefit offset; every krone earned above DKK 25 (€ 3.35) per hour results in a one-to-one reduction in benefits. To counteract these limited financial incentives to work, the law requires recipients to actively seek employment or participate in training programs, with caseworkers regularly assessing adherence. Still, political concerns remain regarding insufficient work incentives due to high implicit tax rates, these concerns have led to a series of reforms aimed at promoting work.[10]

We study a reform passed into law in March 2016 and implemented in October 2016. The economic context for the reform featured relatively low unemployment of around 4% and a steady GDP growth rate of around 2%. The reform impacted all welfare recipients and had two key components: it set a monthly upper limit on total transfers received while on welfare (the sum of welfare benefits and housing support) and it introduced a work requirement of at least 225 hours in the last 12 months to remain eligible for the full level of benefits.

---

[8] At the time of the 2016 reform, the asset threshold was DKK 10,000 (≈ USD 1,500) for individuals and DKK 20,000 (≈ USD 3,000) for couples. The household income limit under the mutual support obligation ('gensidig forsørgerpligt', a term for the legally defined financial requirements of Danish spouses) was DKK 14,575 (≈ USD 2,150) if children were present in the household, and DKK 10,968 (≈$1,600) otherwise. Welfare recepients were allowed to retain the first DKK 25 (≈ USD 3.60) earned per working hour without a reduction in welfare benefits. Earnings above this amount, however, resulted in a corresponding decrease in welfare, effectively offsetting additional income. See "Vejledning om satser mv. 2016".

[9] Special support ("særlig støtte") is a supplementary, tax-free entitlement designed to help welfare recipients with exceptionally high housing costs or significant family obligations. It covers housing expenses exceeding a defined threshold after general housing assistance has been applied and is conditional on the absence of a reasonable, cheaper housing alternative. In practice, however, special support plays a limited financial role: among our population of interest, recipients average only about €9 per month, compared to €356 from general housing assistance and €1,864 from cash benefits.

[10] See, for example, the Danish Economic Council (2015). The disincentives for low-income families to achieve self-sufficiency due to high effective marginal taxes on wage gains are a recurring theme in most welfare states. For recent US examples, see Ilin and Sanchez (2023), where the phaseout of transfer programs leaves single-parent families with no financial gain from earnings increases between $11,000 and $65,000 of earned income. To appreciate the perennial nature of this issue, consult, e.g., Anderson (1978).

In practice, the upper limit was set such that only individuals who received regular welfare benefits and some housing support were at risk of facing the upper limit. The upper limit on total benefits was predicted to cause a non-negligible decrease in benefits received absent any adjustments in hours worked. As shown in Table A1, a typical single parent on welfare with two children was expected to have a disposable income of DKK 13,100 (€1,756) before the reform but only DKK 10,500 (€1,408) after the reform, a reduction of almost 20%. As discussed above, the reform led to a substantial reduction in realized cash benefits. The reduction in benefits made staying on welfare less attractive, but the upper limit on total transfers also created a more subtle incentive to work for those affected. It effectively nullified the benefit offset for individuals with a low number of hours worked by keeping total transfers constant. In other words, any reduction in welfare benefits due to taking up of work was offset by an equivalent increase in housing support.

Recipients had to work 225 hours (about six weeks of full-time work) over the previous 12 months to maintain their benefits. Only regular working hours counted; subsidized employment did not. Compliance was assessed monthly using a rolling 12-month window. When the legislation took effect in October 2016, the requirement was initially set at 113 hours, corresponding to the six months (April-September) from the passage of the law to its implementation. From October 2016 to March 2017, the hours requirement gradually increased, with the full 225-hour requirement coming into effect in March 2017.

The policy's strictness varied considerably with marital status.[11] For couples, if one partner did not meet the work requirements, their individual welfare benefits were withdrawn. If both partners failed to meet the requirements, benefits for one individual were withdrawn. Once both fullfil their work requirements, both receive full benefits again. For singles, failure to meet the requirement resulted in a benefit reduction of DKK 1,000 per month.

Municipalities had some leeway in applying the 225-hour work requirement. Caseworkers could exempt welfare recipients they felt had a limited ability to engage in a gainful activity in a given month from the requirement. By the summer of 2016, the Danish Minstry of Employment estimated that roughly 50% of all welfare claimants were exempt from the hours requirement (though not from the upper limit on total transfers).[12] Unfortunately, our data do not reveal which beneficiaries did or not receive an exemption. In light of these many exemptions, one can (somewhat loosely) view our impact estimates as "intention to treat" effects.

[11] The Danish welfare system treats cohabiting unmarried couples the same as married couples, as do we in our analysis unless otherwise noted.

## 4. *Empirical strategy*

We study the consequences of the welfare reform for mothers' and ultimately children's outcomes. To learn about the full effects of the reform on household resources, we also analyze effects on partner outcomes. We start with the population of mothers who received welfare in March 2016, corresponding to the time at which the reform was passed into law. We align our analysis with the passage of the reform, rather than its implementation, to minimize issues with anticipatory behavior. We view our strategy as conservative because some individuals leave welfare in the interval between passage and implementation who would have done so even absent the reform.

To learn about the effects of the reform, we exploit variation in outcomes around the introduction of the reform in a comparative event study approach that explicitly allows the effects to vary with the temporal distance to the reform. The basic idea is to compare outcomes for individuals in the reform group (mothers or their children depending on the outcome under study) in a given period with their own outcomes immediately before the reform (i.e. before March 2016). However, as shown in Figure 1 and documented further below, welfare participation exhibits clear within-calendar-year dynamics unrelated to the reform. These partly reflect regular seasonal variation in employment but also result from the definition of our population of interest. Requiring receipt of welfare in March 2016 for inclusion in the analysis creates an (inverse) version of Ashenfelter's dip among those analyzed; see Devine and Heckman (1996) and Heckman and Smith (1999) for related discussions. To account for both the seasonality and the outcome dynamics implied by our population restrictions, we create a comparison group consisting of the population of individuals on welfare exactly one year prior, in March 2015. In this comparison group, we compare outcomes immediately *before* March 2015 with outcomes in other time periods. Note that by construction our set-up avoids the comparison group contamination issues discussed in, e.g., de Chaisemartin and D'Haultfœuille (2020).

[12] See Response from the Minister for Employment to the Employment Committee of the Danish Parliament, question no. 480 in the 2015-16 session of the Danish Parliament (https://www.ft.dk/samling/20151/almdel/beu/spm/480/index.htm)

For each observation in the reform group, we denote the time at which the reform was passed into law (i.e., event time) by $t = 0$, and index all periods relative to that point in time. For the comparison group, event time $t = 0$ indicates calendar time one year earlier. We start our analysis well ahead of the reform, with the exact starting point depending on data availability for each outcome. We never go more than 15 months back to avoid interference from a January 2014 welfare reform that primarily targeted the youngest welfare recipients (< 30 years old).[13] We continue our analysis until a year after the passage of the reform, at which point the comparison group gets exposed to reform. This limited follow-up period represents the implicit cost of what we see as a very compelling identification strategy.

Our preferred baseline specification considers a balanced panel of individuals who we observe in a period before and after the reform. Our main estimating equation is the following:

$$Y_{it} = \alpha + \beta \cdot reformpop_i + \sum_{j \neq -1} \delta_j I[j = t] + \sum_{j \neq -1} \gamma_j I[j = t] \cdot reformpop_i + \varepsilon_{it} \quad (1)$$

where $Y$ denotes the outcome of interest, $\delta_j$ are event time dummies, and $reformpop$ indicates that an observation belongs to the cohort exposed to the reform. $\varepsilon$ is an error term, $i$ indicates observations in either the reform or comparison group, and $t$ indicates relative time as described above. The $\{\gamma_j\}$ constitute the parameters of interest; they represent "the effects" of the reform in the population of welfare participants.[14] Note that with individual level panel data, (1) essentially corresponds to a fixed effects analysis with time varying effects of the reform; see e.g. Blundell and Costa Dias (2009) and Lechner (2011). As pointed out by Lechner (2011), estimation using the linear regression framework in (1) without covariates is fully nonparametric. In practice, because many welfare spells last more than a year, some individuals appear in both the reform and comparison groups and so correspond to two observations in the analysis data. To account for this, our main results include standard errors clustered at the individual level.

The key identifying assumptions associated with our approach are 1) no anticipation prior to passage of the reform, 2) parallel trends in outcomes in the absence of the reform, and 3) no other concurrent changes, i.e., no confounding treatments. By anchoring the population prior to the passage of the reform, we limit issues with anticipation. Figure 2 shows that this is a real concern: Google Trends data show that the reform received a lot of attention around the time it passed, with a lesser but not trivial spike in interest in the fall of 2015 due to discussions in the Danish parliament, and another spike around implementation in the fall of 2016.

[13] In principle, both our reform and comparison populations experience the previous reform but that reform relied heavily on changes to caseworker behavior that are likely to occur gradually. For that reason, we might even see some differential pre-trends very early on.
[14] Given our substantive rather than applied econometric interests in this paper we do not obsess about the exact weighted average of individual treatment effects we obtain and instead interpret our estimand as (roughly) the average effect of being on welfare in March 2016 rather than March 2015 for our population of interest.

[FIGURE 2 AROUND HERE]

By anchoring the comparison group to March 2015, we minimize the effects of any differential within-calendar-year outcome dynamics between the two groups. Our specification (1) allows us to directly investigate differences in pre-trends; these will reveal any anticipatory behaviors in the months leading up to the reform.[15] To explore and account for concurrent changes that would lead to confounding bias, we also consider the development over time in outcomes for the non-welfare population in a triple comparative event study.

For child outcomes, we deviate slightly from our preferred specification in (1) that relies on *balanced* individual level panel data and instead consider an unbalanced panel. To illustrate, consider the case of test scores. In the reform group (and analogously in the comparison group), we compare test scores among children who take a test after the reform with test scores among children who take a test prior to reform. There will be considerable overlap in these two populations, of course, but some children who take a test prior to the reform will age out of test-taking after the reform. Conversely, some children will age into the test after the reform. Restricting the population of test takers to be constant in the before-after comparisons would force us to leave out students in the the highest and lowest grades.

## 5. *Data, samples, and descriptive statistics*

### Data sources and outcomes

Our analyses make use of population-wide Danish register-based data. A unique identification number (the central personal register number; CPR) allows us to link individuals across registers and to connect parents to children. We construct a series of outcomes based on these registers. Oftentimes, the underlying data are available at a higher frequency than we can meaningfully analyze, especially for rare events.[16] In these cases, we aggregate to a slightly higher level. Key to our project is, of course, monthly information about welfare participation, benefit payments, and labor market outcomes such as earnings and hours worked, which we draw from the National Income Register.[17]

[15] A standard approach to handling deviations from parallel trends relies on adding conditioning variables, with the underlying assumption then one of conditionoal parallel trends (or conditional bias stability, in some literatures). As discussed by Lechner (2011), including covariates in a standard parameteric linear differences-in-differences setup raises some important issues. Heckman et al. (1998) propose reweighting via matching as an alternative to linear conditioning. Recent work by Sant'Anna and Zhao (2020) replaces the matching with inverse propensity weighting.

In terms of distinct child outcomes, we analyze outcomes that reflect child cognitive and noncognitive skills (e.g. Carneiro and Heckman, 2003) as well as the quality of the home environment. We specifically focus on outcomes that we believe are malleable and indicative of the current situation but also likely to predict future success. We start out by considering children's academic outcomes. In practice, these reflect a combination of the child's underlying ability, along with their own effort and that of their parents (to read aloud, instruct, monitor, and so on). We base these on standardized versions of the nationally administered performance tests in Danish reading and math that have been shown to correlate positively with future academic success (Beuchert and Nandrup, 2018).[18] In practice, the national tests are computerized, self-scoring, and adaptive, meaning that the difficulty of the questions adjusts to the quality of previous answers. Adaptive testing improves the ability of tests to differentiate students in the tails of the distribution (Weiss, 2004). The tests are administered each spring in primary and lower secondary public schools starting from grade 2. Danish reading is tested in grades 2, 4, 6, and 8, while math is tested in grades 3 and 6. See Beuchert and Nandrup (2018) for more information on the specifics of the tests. We might expect the reform to negatively affect test scores either via increasing the stress level in the student's home or by reducing purchases of complementary educational inputs.

Like Dustmann et al. (2024), we study indicators of children's well-being based on the nationally administered school-based individual well-being surveys developed by the Danish Ministry of Education. These surveys are collected in the spring of each year with the purpose of learning about classroom and school-level well-being. Crucially, neither school management nor individual teachers have access to the student-level responses. In our analyses, we focus on the social well-being scale collected for children enrolled in public schools in grades 4-9. The responses to all questions consist of Likert scales with five options, which we coded to range from one to five, with five being the most positive. In practice, we calculate the average response across all questions for each child and our well-being measure therefore ranges from 0-5.[19,20] As with the national tests, social well-being has been shown to correlate positively with academic performance, though the estimates are not very large; see Larsen et al. (2020).

[16] Truly a first world researcher problem.
[17] We worry about the possibility of off-the-books earnings both in the levels sense, which would lead us to understate the resources available to the population we study, and in the treatment effects sense, as the benefit cuts may inspire additional off-the-books activity. To the extent that they do, our estimates of the decline in household resources become too big, with the implication that our treatment effects represent lower bounds on the change in behavior per unit of household resources. The estimates of the incidence of off-the-books earnings for Denmark from the Eurostat Labour Force Survey reported in Statistics Denmark (2018) seem far too low to us, a concern reinforced by the survey's low-single-digit response rate.
[18] Beuchert and Nandrup (2018) also document that the results from the national tests correlate with socio-economic status (SES) in the expected way.

Our final main outcome consists of reports to child protective services because of concerns for the child in question. Child maltreatment has consistently been shown to correlate with (or even cause) future risky behaviors (e.g., Gilbert et al., 2009a; Currie and Tekin, 2012). The data from child protective services include the date of the report, the reason for the concern, and the type of informant. Anyone can express concern, and the report can be anonymous, though the most common source is the school, as we document in Section 6.2. This is a serious measure; within 24 hours, the authorities must not only register the report centrally but also evaluate the need for an urgent response. Depending on the nature of the report, this may or may not lead to an interview with the child ("en børnesamtale"). Child protective services must also inform any reporting professionals about the outcome of the investigation and the support measures taken.[21]

We use absence data from public schools, provided by the Danish Ministry of Education, as well as information about child injuries from hospital admissions data as supplementary outcomes related to

---

[19] The social well-being scale for children enrolled in grades 4-9 consists of the following ten items:

a) How well do you like your school?
b) How well do you like the other children in your classroom?
c) Do you feel lonely?
d) Are you afraid of being ridiculed at school?
e) Do you feel safe at school?
f) Since the start of the school year, did anyone bully you?
g) I feel I belong at my school.
h) I like the breaks at school.
i) Most of the pupils in my classroom are kind and helpful.
j) Other pupils accept me as I am.

The responses to all questions are coded to range from one to five, with five being the most positive. For positive questions like "Do you feel safe at school?" the value five is equivalent to "very often". For negative questions like "Do you feel lonely?" five means "never". In this sense, five is always the best outcome.

[20] Of course, we regret treating ordinal responses as cardinal, even though (almost) everyone else does too.

[21] https://ast.dk/born-familie/hvad-handler-din-klage-om/underretninger#:~:text=Hvad%20skal%20en%20kommune%20g%C3%B8re,for%20barnet%20eller%20den%20unge. Accessed October 3, 2024.

behaviors that could lead to reports to child protective services. For both adults and children, we match our outcomes data to rich demographic information from various administrative registers.

**Samples and descriptive statistics**

From the National Income Register, we first select the 33,960 mothers of children aged 0-18 who were on welfare in March 2016. From this group, we exclude recent immigrants because of a concurrent reform affecting this group.[22] We also exclude mothers not at risk of facing the upper limit on total benefits. As explained above, to be at risk, one would have to receive both welfare benefits and some housing support.[23] Our final 2016 sample (the "reform group") consists of 18,578 mothers and 36,820 biological children.[24] Table A3 describes our sample loss journey in detail.

We subsequently select the corresponding set of mothers on welfare in March 2015 (our "comparison group") as well as their biological children aged 0-18. Importantly, we impose the exact same exclusion criteria as for the reform group. There are 22,055 unique mothers in the two groups. 17,254 mothers appear in both groups and 4,801 appear in only one the reform group or only the comparison group. Table 1 shows that mothers on welfare in March of 2015 and 2016 resemble each other closely in terms of a wide range of observed characteristics. To grasp the potential importance of the 225 hours requirement for the population under study, note that mothers on welfare worked on average three hours in March. Accordingly, working 225 hours per year would constitute a substantial change.

[TABLE 1 AROUND HERE]

Table 1 also shows how our reform and comparison groups compare to the overall population of adult Danish mothers not on welfare and clearly documents that the former group is severely disadvantaged in terms of background characteristics and attachment to the labor market. The mothers in our analysis are less likely to be married and to cohabit, have lower educational attainment, and have much less work experience. Except for dental care, they also interact much more with health care professionals and are much more likely to be victims of or to commit crime. Appendix tables A4 and A5 show similar patterns for the biological fathers and for the current partners of the mothers.

Table 2 describes the children of the mothers in our analysis and compares them to the corresponding overall population. As with the mothers, the children of the reform and comparison group mothers have very similar observed charactteristics. Children of mothers on welfare are clearly disadvantaged compared to other children and this is obvious already from birth: their mothers were more likely to be overweight and to smoke, and their parents were more likely to be teenagers at the time of birth. They are also more likely to be enrolled in special schools and have vastly lower test performance. Moreoever, their risk of having at least one report to child protective services is six times higher than for a child in the overall population.

[22] Those not living in Denmark for at least seven out of the last eight years.
[23] Table A2 compares the sample at risk of facing the upper limit with other mothers on welfare.
[24] We do not include adoptive or foster children. We do include biological children even when not co-resident.

[TABLE 2 AROUND HERE]

## *6. Consequences of welfare reform*

This section shows our main empirical findings. Since the overall purpose of the reform was to incentivize labor market participation, and ultimately lower levels of welfare participation, we start by exploring these margins for the mothers. Equipped with those findings, we then move to explore the consequences for child academic outcomes and well-being. We end the section with a range of robustness analyses and an investigation of subgroup heterogeniety in impacts.

### *6.1 Effects on mothers' outcomes*

Results on mothers' hours worked appear in Figure 3. The upper panel shows the differences in hours worked over time for our reform and comparison groups. The lower panel shows the estimated effects of the reform from the comparative event study estimation (1) anchored in February just prior to the passage of the reform. Note first that the estimated effects of the reform are essentially zero in the months prior to its passage and all estimates are statistically insignificant too. This reassures us that our estimation approach actually manages to balance pre-trends and that welfare participants did not change their behavior in anticipation of the reform before it passed into law. In an absolute sense, we estimate small, positive effects on hours worked as a consequence of the reform. In March 2017, for example, the estimated effect equals four hours. In a relative sense, effects are large: four hours correspond to as much as 35% of the comparison mean at that point in time. Still, to put this into perspective, only 13% of mothers on welfare benefits in March 2016 managed to accumulate at least 225 hours in the 12 months from April 2016.

[FIGURE 3 AROUND HERE]

To learn more about individuals' response margins (i.e., about the heterogeneity of the reforms effects on labor supply), we have also explored the extent to which the reform affected the propensity to work at all (any hours in a given month); to work part time (at least 80 hours per month); and to work full time (at least 160 hours per month). Results are shown in Table A7. We find that the reform mostly affect the tendency to work at all (about 5 percentage points more in March 2017 compared to a mean in the comparison group of 10%) but also had some effect on part time work (an increase of 2 percentage points compared to a mean in the comparison group of 7%). The reform had no impact on full time work. Table A8 shows that mothers typically worked in low-skilled jobs, for example as aides in elderly care (17% of mothers with some employment in 2016); as cleaning assistants (12%); and as pedagogical assistants (9%).

Figure 4 considers mothers' welfare participation. Pre-trends are well-aligned for the nine months (August and onwards) prior to the reform. We suspect that the small negative estimates of roughly 1 percentage point in the months long before the passage of the reform are driven by the implementation of the January 2014 reform discussed above. After the reform–from July and onwards, and particularly from October when the reform was fully implemented–we detect negative and statistically significant effects on welfare participation. Again, the effects are modest in an absolute sense: the estimate in March 2017 is -0.082, corresponding to 10% of the comparison group mean.

In short, some but not many mothers in this population managed to leave welfare entirely and only a small fraction increased their labor supply to an exent sufficient to avoid potential monetary sanctions associated with the 225 hours requirement. The timing of the responses suggests to us that mothers in this population are rational and forward looking but, since there were no anticipatory responses, either derive substantial disutility from work or are just not well-informed about upcoming changes to current policies. Recalling the relatively low aggregate unemployment at the time, we do not put much weight on search frictions in explaining the temporal patterns.

Figure A1 shows that the reform also affected the mothers' partners (of whom about half were also on welfare in March, just prior to the reform). As for the mothers, we estimate small, positive effects on hours worked of around six and small, negative effects on welfare participation of roughly six percentage points one year after the passage of the reform. The combination of the reduction in benefits through the upper limit, the risk of monetary sanctions, and the lack of any substantial increases in earnings through hours worked *de facto* meant that most families had fewer resources available after the reform. As described in Table A9, we estimate that one year after the passage of the reform, monthly household discretionary household was on average lower by just above €300 because of the reform.[25,26]

[FIGURE 4 AROUND HERE]

### *6.2 Effects on child outcomes*

Given insights into how the reform affected mothers' labor supply and family resources, specifically resulting in lower discretionary income but slightly more hours worked, we move on to investigate effects on child outcomes.

Table 3 first shows the results for academic performance as measured by the national test scores in Danish reading and math.[27] In parallel to the adults, we consider children born to mothers who were on welfare in March of 2016 (2015) to be the treated (comparison) cohort. Importantly, we detect no negative effects on test scores as a consequence of the reform; if anything, our results actually indicate small but statistically insignificant upticks in academic performance corresponding to 1.5% of a standard deviation in Danish reading and 3.1% in math (both less, in absolute value, than the corresponding estimates for the "pre" period). It is possible, of course, that we fail to estimate strong results since test scores reflect skills that accumulate over time while we estimate only the immediate effect of the reform.[28] At the same time, the lack of a short-run response on test scores leads us to downplay to some extent reform effects operating through household stress and disruption, as we would expect that channel to reduce test performance conditional on underlying achievement.

[TABLE 3 AROUND HERE]

To probe into contemporaneous consequences for children's state of mind, we next explore self-reported social well-being, constructed from the nationally administered well-being surveys developed by the Danish Ministry of Education. The survey is gathered each spring, which implies that we consider the spring of 2016 (2015) as the pre-period for the reform (comparison) population. Social well-being is our only measure that relies entirely on children's own reports. It informs us about how they perceive themselves to be thriving (or not), keeping in mind that the questions primarily concern children's well-being at school, not at home. Results appear in Table 4. In addition to our baseline specification in column (1), we include a specification (2) that allows for possible calendar time trends in (the measurement of) overall well-being. We detect negative effects on well-being of 0.03 and 0.06 percentage points in the two specifications, corresponding to 4% or 8% of a standard deviation or about 50% of the difference between the population of children on welfare and other children documented in Table 2.[29,30]

[25] As noted above, we define discretionary income as the sum of mother's and any partner's earnings and transfers minus taxes and rent.

[26] In addition to the outcomes described here, we also considered primary care visits, visits to private practicing psychologists or psychiatrists, and emergency department visits, all for mothers. We found no meaningful evidence of effects.

[27] We tested whether the reform affected test-taking for the national tests and survey-taking for the social well-being measure. We did not find evidence of this. 91.8% (91.3%) of children in the reform (comparison) population take the national test in Danish reading in 2016. 87% of children in both the reform and comparison population answer the social well-being survey in 2016.

[28] Landersø et al. (2020) do find that 9th grade test scores are sensitive to the school start age of younger siblings; another type of stressor. The authors argue that this is likely because delaying the school start of a younger sibling allows parents to redirect resources towards the dimensions in older siblings' upcoming exams that are most easily improved.

[TABLE 4 AROUND HERE]

As noted above, the school management and teachers do not have access to the individual level responses, yet children may still worry that their reports will reflect negatively on their parents and this loyalty conflict may impact their responses.[31] In the US, for example, fewer than one percent of the reports to child protection services are made by the victims themselves, indicating that children are reluctant to reach out to authorities about serious family issues; see Gilbert et al., (2009b). If children worry about bringing issues at home to the attention of professionals, analyses of children's well-being that relate to changes in their parents' circumstances cannot only rely on children's self-reports. Partly for this reason but also to explore effects on a more severe outcome indicative of children's well-being and the quality of the home environment, we analyze reports to child protective services because of concerns for the child in question.

The upper panel of Figure 5 reveals that children in our sample face a quarterly risk of having at least one report made to child protective services of between seven and eight percent.[32] Table A11 indicates that the most common reasons for concerns in our population are child externalizing behaviors (22% of all reports); insufficient care from parents (16%)[33]; and high levels of conflict in the family (13%). Informants are primarily school staff (21%) or health care providers (13%), though anonymous informants also play a role (8%).

[29] The raw data show a clear trend following the introduction of the student survey in 2014. For this reason, we assign a larger weight to the specification in column (2), though of course the two estimates tell a common story in terms of sign, statistical significance, and rough magnitude.
[30] Appendix Table A10 shows the results for each of the sub-questions included in the social well-being score. We observe small effects across all questions.
[31] This hypothesis was also brought up in personal communication with the previous chair of the Social Workers' Union Majbrit Berlau (January 12, 2021).
[32] The number of reports to child protection services have generally increased over the period that we study. This is most likely because of early reforms ("The Child's Reform" (Barnets Reform) from 2011 and "The Abuse Reform" (Overgrebspakken) from 2013). Our calculations show that increases in the number of reports related to particular children over time drives the overall increase, rather than an increase in the number of children ever subject to a report.

The impact estimates appear in the lower panel of Figure 5. We follow our approach from above and consider the first quarter of 2016 (2015) as the pre-period for the reform (comparison) population. Note that because the gathering of this particular data source starts in 2015, we are unable to explore pretrends. We find substantively meaningful and statistically significant increases in reports to child protective services. Though not especially large in an absolute sense (just 0.8 percentage points in the quarter of the reform and 1.9 percentage points in the quarter following the reform) they are substantial in a relative sense (10% of the comparison mean in the quarter of the reform; 26% in the quarter following the reform).

Our results indicate that the reform increased the likelihood that children come to the attention of professionals. This can either be because of child behaviors that lead professionals, such as their teachers, to react and file a report, or because increased risks at home get directly detected by individuals who engage with the child's family. Our results corroborate those of Kovski et al. (2022) who document links between the presence and generosity of EITC and state-level rates of child maltreatment in a standard panel data exercise.[34]

[FIGURE 5 AROUND HERE]

*6.3 The role of concurrent changes: a potential threat to the validity of the design*

One potentially important concern relates to changes in context across the various outcome domains that coincide with (or postdate) the reform and thus may confound our findings. Such changes could, for example, include shocks to labor demand, epidemics, or even revisions of the institutional setting. To explore the sensitivity of our findings to such factors, we implement a triple comparative event study that exploits a population that was unaffected by the reform. We track the outcomes for this population during the exact same period as the reform cohort, allowing us to estimate and account for overall trends. We foreground the standard comparative event study implicit in (1) relative to the triple comparative event study because we prefer the simplicity and transparency of the former.[35]

[33] Insufficient care, or neglect, relates to factors such as physical care (lack of food, clothes, and protection), emotional care (love and affection), developmental care (basic educational needs), and access to appropriate medical care. See Gilbert et al. (2009b) and the Danish national guidelines for medical staff here: https://www.sundhed.dk/borger/patienthaandbogen/boern/sygdomme/socialpaediatri/omsorgssvigt-og-overgreb-mod-boern-og-unge-diagnostik/.

[34] In addition to the outcomes reviewed here, we also investigated labor supply effect and crime for youth ages 15-18 and school changing for children of school age. We found nothing worthy of space in the text.

Specifically, we propose to use the populations of mothers who were *not* on welfare in March 2016 (and their children) to capture concurrent events for the reform sample and the population not on welfare in March 2015 to capture trends for the comparison sample. To investigate whether this population was affected by the reform, we first explore whether the reform affected inflows into welfare for mothers. Figure A2 shows the share of all mothers on welfare benefits during 2015 and 2016 as well as the difference between these shares. The share on welfare is essentially flat throughout 2016 and the difference between the shares across the two years is also close to constant. We use mothers not on welfare for this exercise, rather than mothers on welfare but not at risk of reaching the benefit cap, because this latter group still experiences the work requirement part of the reform.

We present the results for the outcomes from above based on the triple comparative event study for mothers' outcomes in Figure 6. Panel A shows the results for hours worked, while Panel B depicts those for welfare participation. Fortunately, we do not uncover any evidence that other concurrent events drive our findings.

[FIGURE 6 AROUND HERE]

Table 5 displays estimates from the triple comparative event study for children's outcomes. The estimated effects on Danish reading, math, and the indicator for reports to child protectives services turn out very similar to those from the model in (1). The only real difference between the results from our main specification and those from Table 5 is that the estimated effect on social well-being becomes essentially zero when accounting for overall trends.

[TABLE 5 AROUND HERE]

### *6.4 Refined insights: Heterogeneity analyses and alternative mother and child responses*

There are good reasons to think that effects on the children may vary with the mother's civil status. To the extent that a partner can help alleviate any stress incurred by the reform, assist with the children, and contribute additional income, we expect children in households with multiple adults rather than just one to exhibit greater resilience in response to the reform.[36] To take this hypothesis to the data, we estimate separate effects by civil status, distinguishing single parent households from those with a married or cohabiting parent. We also investigate whether child sex acts as a moderator, although the extant literature does not offer a clear prior about whether we would expect larger effects for boys or girls (e.g., Almond and Currie, 2011). Finally, for reports to child protective services, we explore the moderating role of child age, inspired by obvious differences in the experiences of children as they grow.

[35] The late John DiNardo offered the rule of thumb that the plausibility of differences estimators decreases with the square of the number of differences.

As seen in Table 6, the reform increases reports to child protective services more (and earlier on) in single headed households. We also detect non-negligible positive effects on test scores for children in households with more than one adult, with the estimate associated with math being significant at the ten percent level. Surprisingly, we do not find that the effects of the reform on family adjusted discretionary income varied by much whether a partner was present or not.[37]

[TABLE 6 AROUND HERE]

[TABLE 7 AROUND HERE]

We do not find differences across child sex in terms of academic outcomes or social well-being as shown in Table 7.[38] But boys clearly account for the increase in reports to child protective services; they have a larger estimate, and, unlike the girls, we can safely reject the null of a zero effect for the boys.[39] Table A11 shows that boys, who make up 51% of the data, make up 54% of all reports. Boys are substantially more likely to have a report because of crime but, otherwise, the concerns motivating the reports more or less balance across boys and girls. In line with this, informants are more often police in the case of boys. With regards to age, we find larger effects on child protective services reports for children aged 6-14 both compared to preschool children, who are generally less visible to outsiders, and (less obviously) compared to the older children, who have more independence and so can more easily escape from high levels of conflict at home. We caution against over-intepretation of all our subgroup estimates as, while the point estimates often meaningfully differ, our data rarely allow us to reject the null of equal effects.

[36] Kearney (2023) makes this argument at length.

[37] Yet with mean family size adjusted discretionary income (in the comparison group one year after the passage of the reform) of €1,545 for single headed households and €1,845 for other households, the reform did affect the former group porportionately more.

[38] The register data record sex at birth rather than self-identified gender.

[39] The share of reports to child protective services as well as the types of concerns expressed by informants are similar across boys and girls.

We now unpack our results in another way by studying potential mediating behaviors. First, the reform might lead mothers to move their families to less expensive housing, which could, in itself, cause (perhaps transitory) disturbances to child and family well-being. We use monthly information about place of residence to explore the effects of the reform on the propensity to move. Table A9 shows that the reform increased the probability of a move to another address by just over one percentage point one year after passage of the reform. Despite the fixed costs associated with changes in residential location, this population moves a fair amount; in the comparison group, as many as 17% of mothers had moved one year after the passage of the reform. Thus, while the reform did induce some families to move, we think it is unlikely to fully account for the effects described above.

Second, the reform might lead to an increase in school absences, due to stress in the family or less parental bandwidth for monitoring child behavior. While interesting in its own right, this outcome also matters because 6% of the reports to child protective services result from high levels of school absence. We use data on absences for students enrolled in public schools provided by the Ministry of Education. Looking again to Table A9, we find statistically significant upticks in absence rates due to the reform in the non-summer quarters; in the fourth quarter after the passing of the reform, we estimate an increase in children's absence rate of 0.007 relative to mean monthly absence rates of around 0.09. We think of this result as corroborating our main findings.

Third, the reform might lead children to experience more injuries, perhaps because children act out more or because financial stress directs parents' attention elsewhere. As discussed above, the reports to child protective services were primarily driven by externalizing behaviors in the child; insufficient care from parents, and high levels of conflict in the family. We explore the effects of the reform on the incidence of injuries by means of diagnosis data from hospital admissions (in- and outpatient).[40] Table A9 shows an increase in the prevalence of injuries of 0.6 percentage points in the second quarter of the reform, where the overall prevalence of injuries equals 5.1% in the comparison population. Note that this quarter contains the summer months when school is out. Table A13 shows that common injuries among children are contusions to the wrist or hand (9%), head wounds (9%), dislocated joints at the ankle or foot level (8%), fractures of the forearm (6%), and dislocated joints at the wrist or hand level (6%). Again, we see this finding as supporting our main conclusions above.

[40] We include the ICD-10 diagnosis codes S00-T88 that cover injury, poisoning and certain other consequences of external causes.

## 7. *Conclusion*

This paper uses register-based data to analyze the consequences of a recent major Danish welfare reform for children's human capital and well-being. In addition to work requirements, the reform introduced an upper limit on welfare benefits that resulted in considerable reductions in welfare payments for some households. We implement a comparative event study that contrasts individuals on welfare at the time of reform announcement before and after the reform with the development in outcomes for the group of individuals on welfare exactly one year prior. Our analysis documents that mothers' propensity to receive welfare decreased only very slightly as a consequence of the reform, just as we observe only a small increase in hours worked. The lack of response to the work requirement allows us to frame the reform as consisting of a straightforward reduction in benefits for some households. Among these households, the reform often implies a considerable decrease in families' discretionary income per household member, especially in single mother households.

We do not detect negative effects on short-run children's academic performance but do, however, document small negative effects on children's self-reported school well-being along with substantial upticks in reports to child protective services for children exposed to the reform. Notably, reports to social services were primarily due to child externalizing behaviors, insufficient care by parents, or high levels of conflict in the family. The increases in reports to child protective services were especially large in single parent households.

Given that the reform did not have large effects on the likelihood of receiving welfare for the adult population, one could reasonably scale our intention-to-treat estimates with the share still receiving welfare to arrive at an average treatment effect for those still receiving welfare. As about 70% of the mothers in our analysis on welfare at the time the reform passed into law remain on welfare one year later, this would increase our impact estimates around that time by about 40%. This rescaling increases our estimated effect on social well-being to 0.08 points or 12% of a standard deviation and that on reports to child protective services to 2.7 percentage points.

Relative to the extant literature, our work adds rich, high quality administrative data that includes outcomes not previously explored in this literature, clean and compelling causal identification, and a welfare reform with fewer moving parts than most (essentially just a decrease in household income), which aids interpretation. While we do not perform a full social cost-benefit analysis of the reform we study, our findings suggest that such analyses of other reforms likely understate the negative side of the ledger if they fail to take account of the sorts of behaviors that lead to increases in reports to child protective services and less happiness at school.

How should we think about externality validity, i.e. about the relevance of our findings to places other than Denmark, with its legos, flexicurity, and its tough but generous welfare state? Carrying our analysis over to other Nordic and central European countries should elicit little controversy given the similarity in institutions. Carrying our analysis onto developed countries with less encompassing welfare states, such as the United States, seems reasonable to us as well. We acknowledge that these countries do less to insure some of life's shocks, but the reform we study here only reduces cash transfers, while leaving the broader safety net context unaffected. For example, viewed as relatively short-term effects of a marginal but meaningful and ongoing change in the household budgets for some disadvantaged households. As such, we would, in the US context, put some weight on them in predicting the near-term effect of modest but not trivial cuts to, say, Temporary Assistance to Needy Families (TANF) benefits, particularly in more generous states.

We close with some limitations. Our compelling identification strategy limits us to but one year of follow-up. The effects we estimate may represent (in whole or in part) transitory responses to the shock of the reform, or they may represent (in whole or in part) the tip of the iceberg of larger long-term effects on related outcomes. We have not investigated equilibrium effects of the sort considered in Lise, Seitz and Smith (2004) or Crépon et al. (2013), wherein policy that presses one subset of the low wage labor market to search more intensely for work has implications for the optimal search behavior of others in the low wage labor market and thereby for the outcomes of individuals outside a partial equilibrium evaluation like ours. The very small hours effects we find in partial equilibrium lead us to have a strong prior that such spillovers will not matter much in our context.

Finally, some readers may worry that we simultaneously explore various outcomes and mechanisms at several points in time, yet do not perform any of the epistemic rituals related to multiple hypothesis testing. Given an existing literature that, to the extent that it considers children at all, focuses almost entirely on a limited set of academic outcomes, we think of our study as a hypothesis-generating exploratory analysis that can inform additional work in the area (Institute for Education Sciences and National Science Foundation, 2013). This matters as many researchers do not have easy access to a broad span of outcomes and will have to choose ex ante which to gather. Our study may serve as a guide for such choices.[41]

[41] We did not undertake a pre-analysis plan. Had we done so, we would have specified labor supply and welfare receipt as the confirmatory outcomes for adults and test scores as the confirmatory outcome for children. For casual Bayesian readers who wish to apply informal multiple testing adjustments, the text describes, or mentions in footnotes, all of the outcome variables we considered.

## Figures and Tables

Figure 1

Monthly sum of government transfers, household equivalent (€)

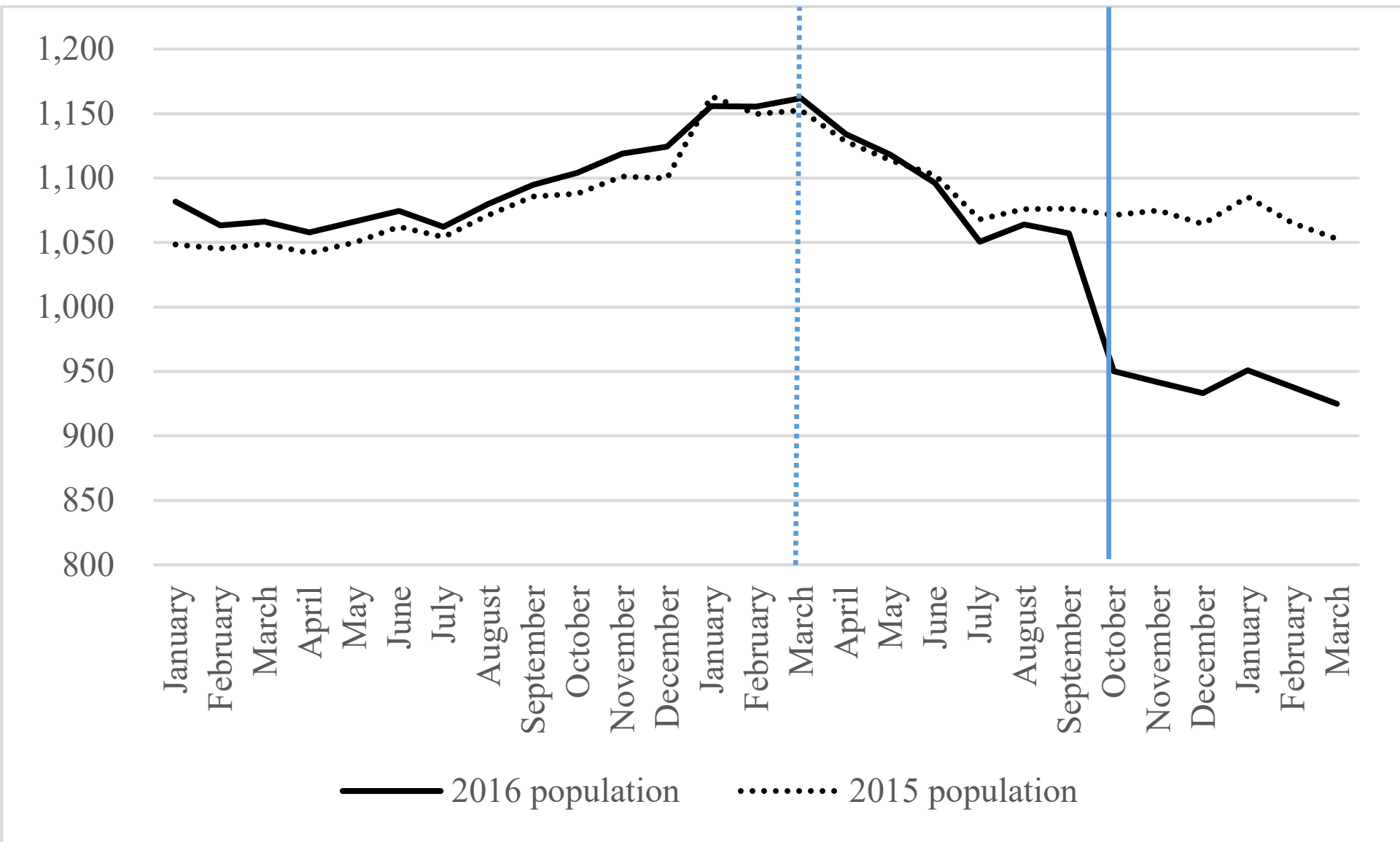

*Notes:* The figure shows amounts of cash benefits paid out to our population of "at risk" moms and their households in 2015 € (deflated using the consumer price index), while accounting for household size by dividing by the square root of family size. 2016 (2015) population consists of "at risk" mothers of children aged 0-18 on welfare in March 2016 (2015). See Section 5 defines the "at risk" population. The dotted vertical line indicates the passage of the reform (March 2016 (2015) for the 2016 (2015) population); the solid vertical line indicates the timing of the reform implementation (October 2016 (2015) for the 2016 (2015) population).

Figure 2

Google Trends data, "Upper limit on welfare benefits" ("Kontanthjælpsloft")

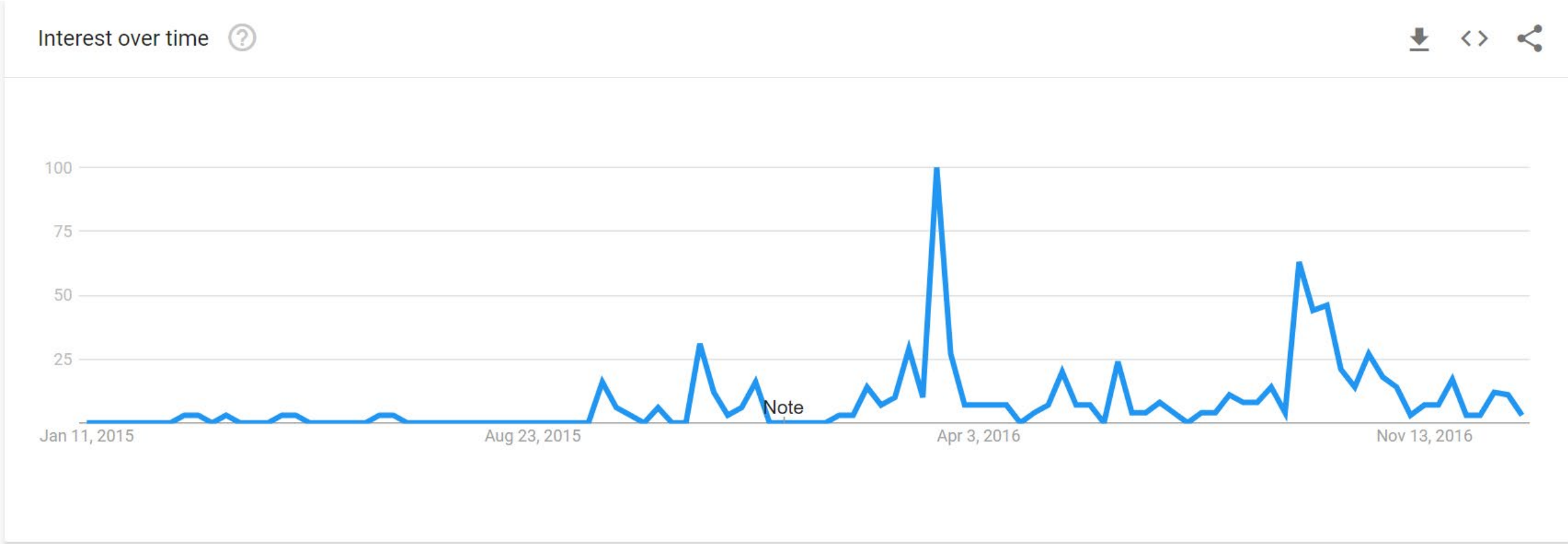

*Notes:* The figure shows Google Trends data for the search term "Upper limit on welfare benefits" ("Kontanthjælpsloft"); see https://support.google.com/trends/answer/4365533?hl=en for details on the approach. The numbers represent search interest relative to the highest point on the chart. A value of 100 is the peak popularity for the term. A value of 50 means that the term is half as popular. A score of 0 means there was not enough data for this search term, where Google Trends leaves "not enough data" undefined.

Figure 3

Effects of the reform on mothers' hours worked

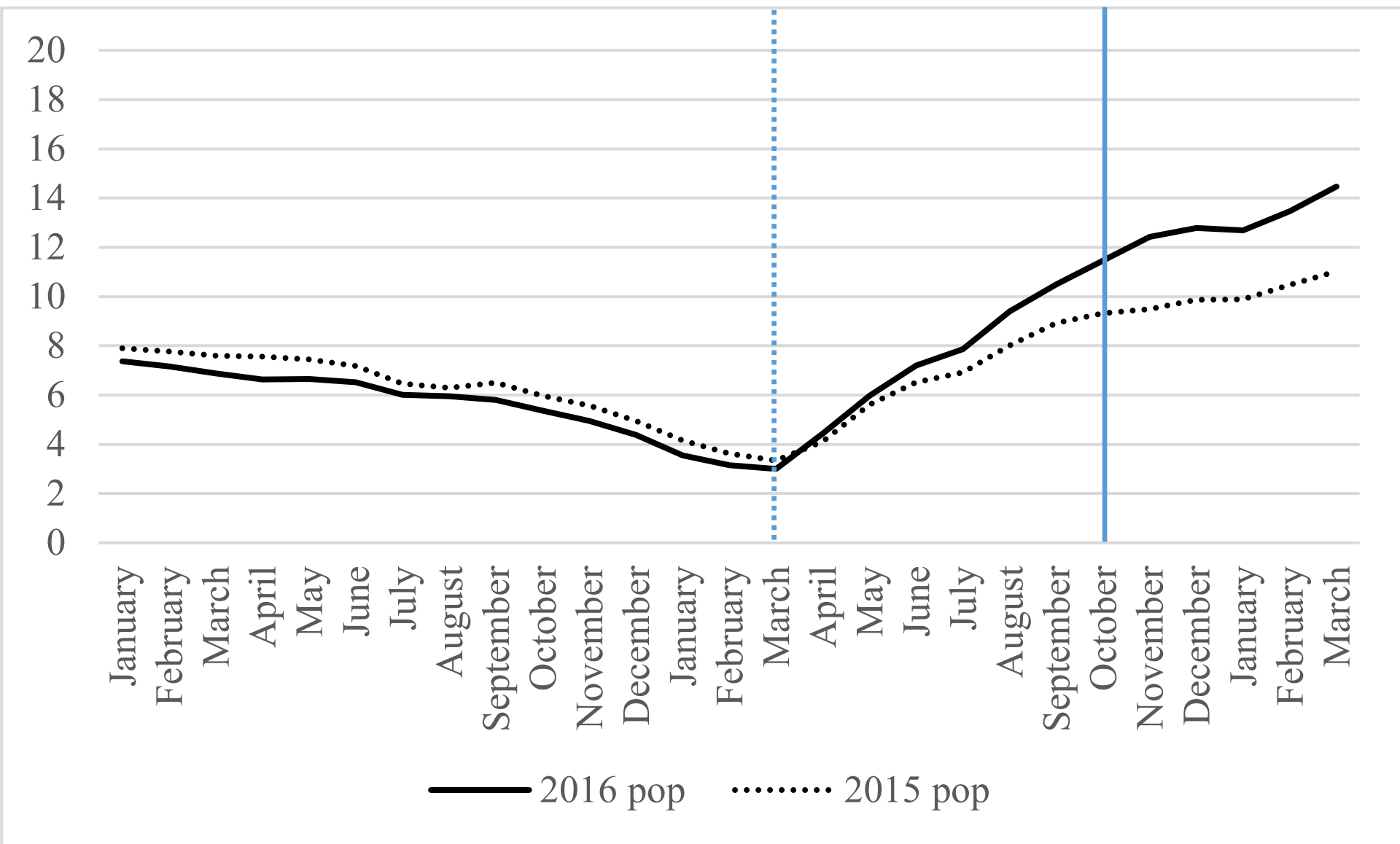

Panel A: Mean hours worked

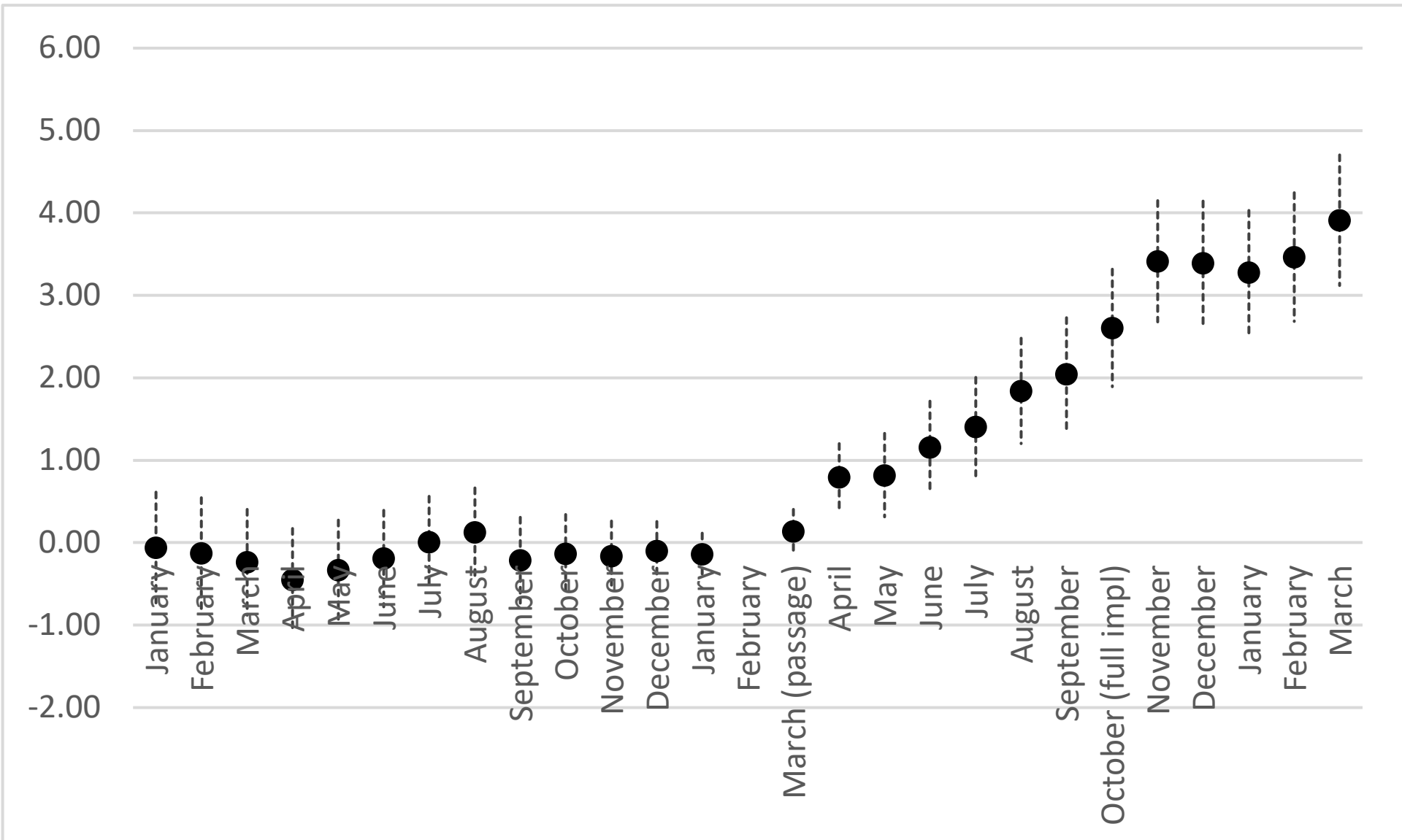

Panel B: Estimated effects of the reform

*Notes:* Panel A shows mean hours worked; Panel B shows the estimates and 95%-confidence intervals from a comparative event study estimation anchored in February, just prior to the passage of the reform. 2016 (2015) population consists of 18,578 (19,716) mothers on welfare benefits in March 2016 (2015). The dotted vertical line in Panel A indicates the passage of the reform (March 2016 (2015) for the 2016 (2015) population); the solid vertical line indicates the timing of the reform implementation (October 2016 (2015) for the 2016 (2015) population). The full set of estimates is shown in Table A6.

Figure 4

Estimated effects of the reform on mothers' welfare participation

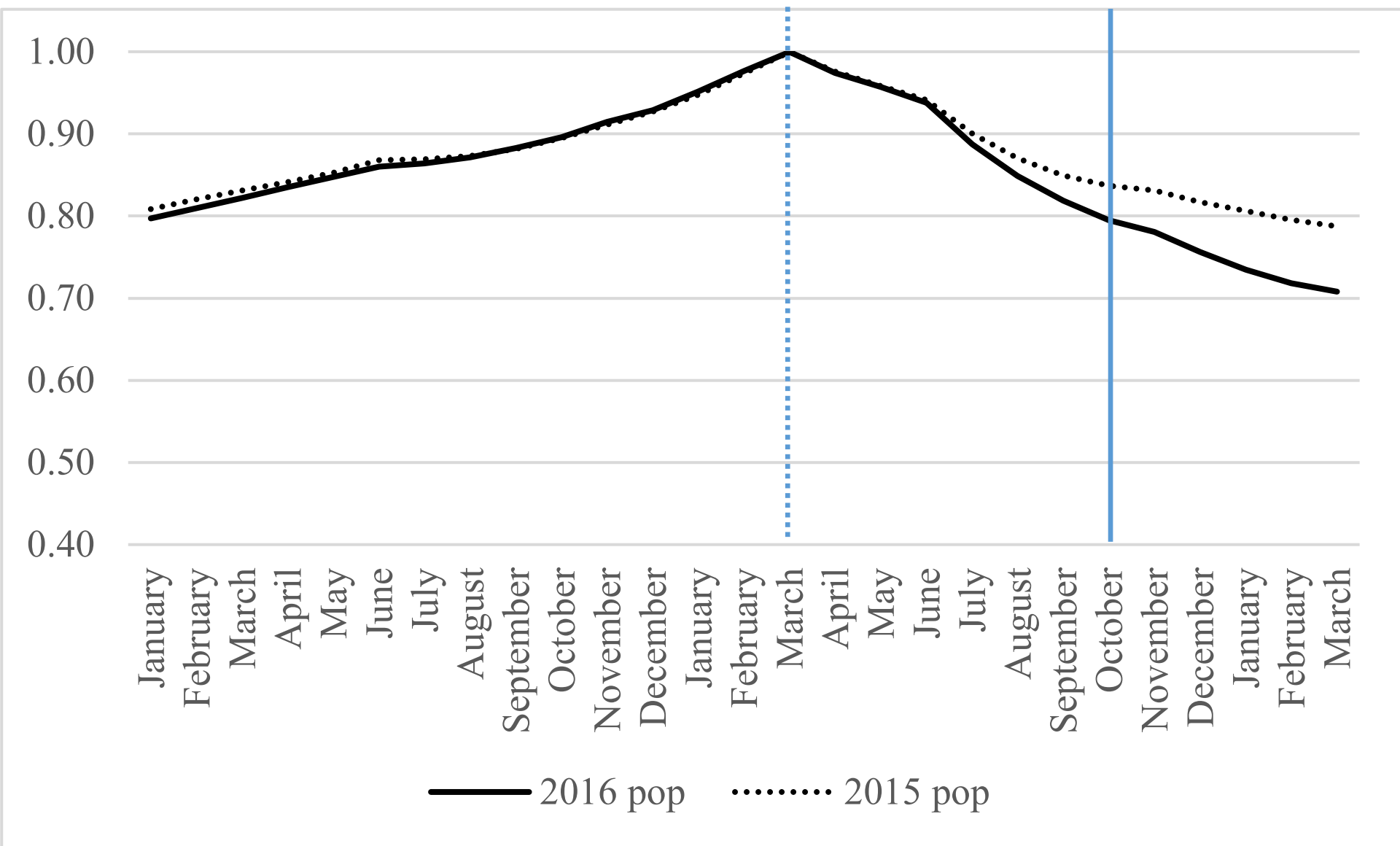

Panel A: Share on welfare

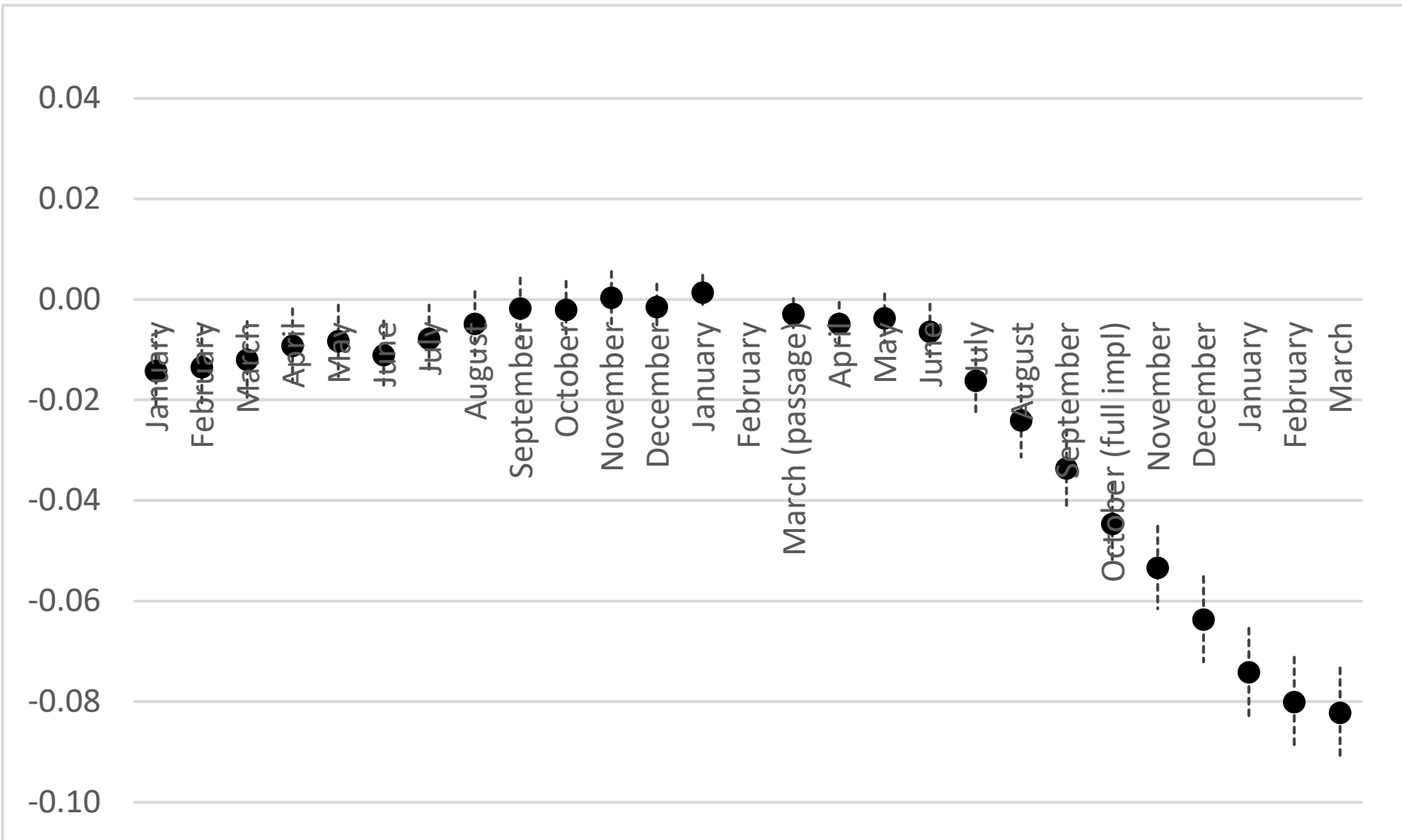

Panel B: Estimated effects of the reform

*Notes:* Panel A shows the share receiving any welfare benefits; Panel B shows the estimates and 95%-confidence intervals from a comparative event study estimation anchored in February, just prior to the reform. 2016 (2015) population consists of 18,578 (19,716) mothers on welfare benefits in March 2016 (2015). The dotted vertical line indicates the passage of the reform (March 2016 (2015) for the 2016 (2015) population); the solid vertical line indicates the timing of the reform implementation (October 2016 (2015) for the 2016 (2015) population). The full set of estimates appear in Table A6.

Figure 5

Estimated effects of the reform on reports to child protective services

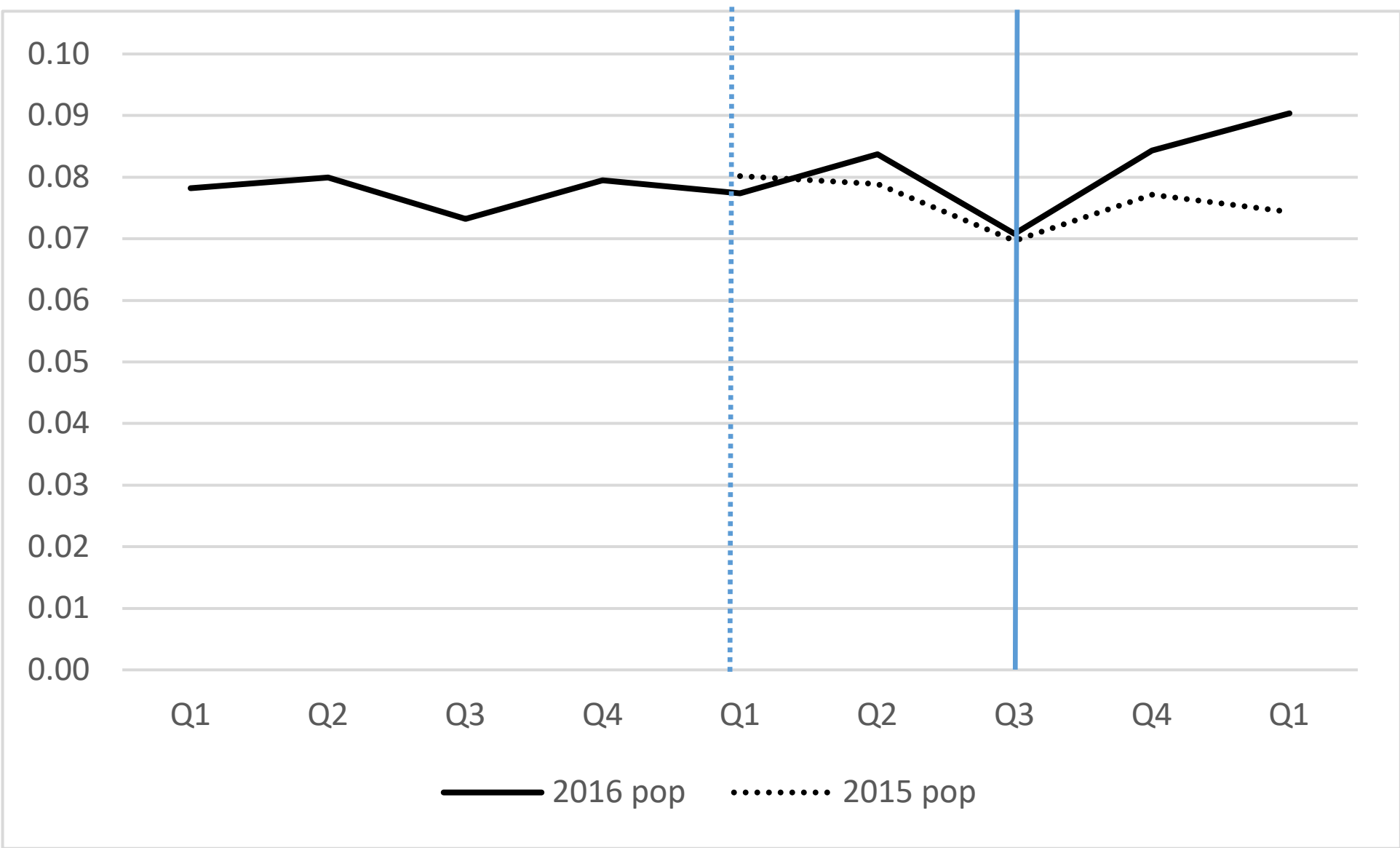

Panel A: Share with at least one report made

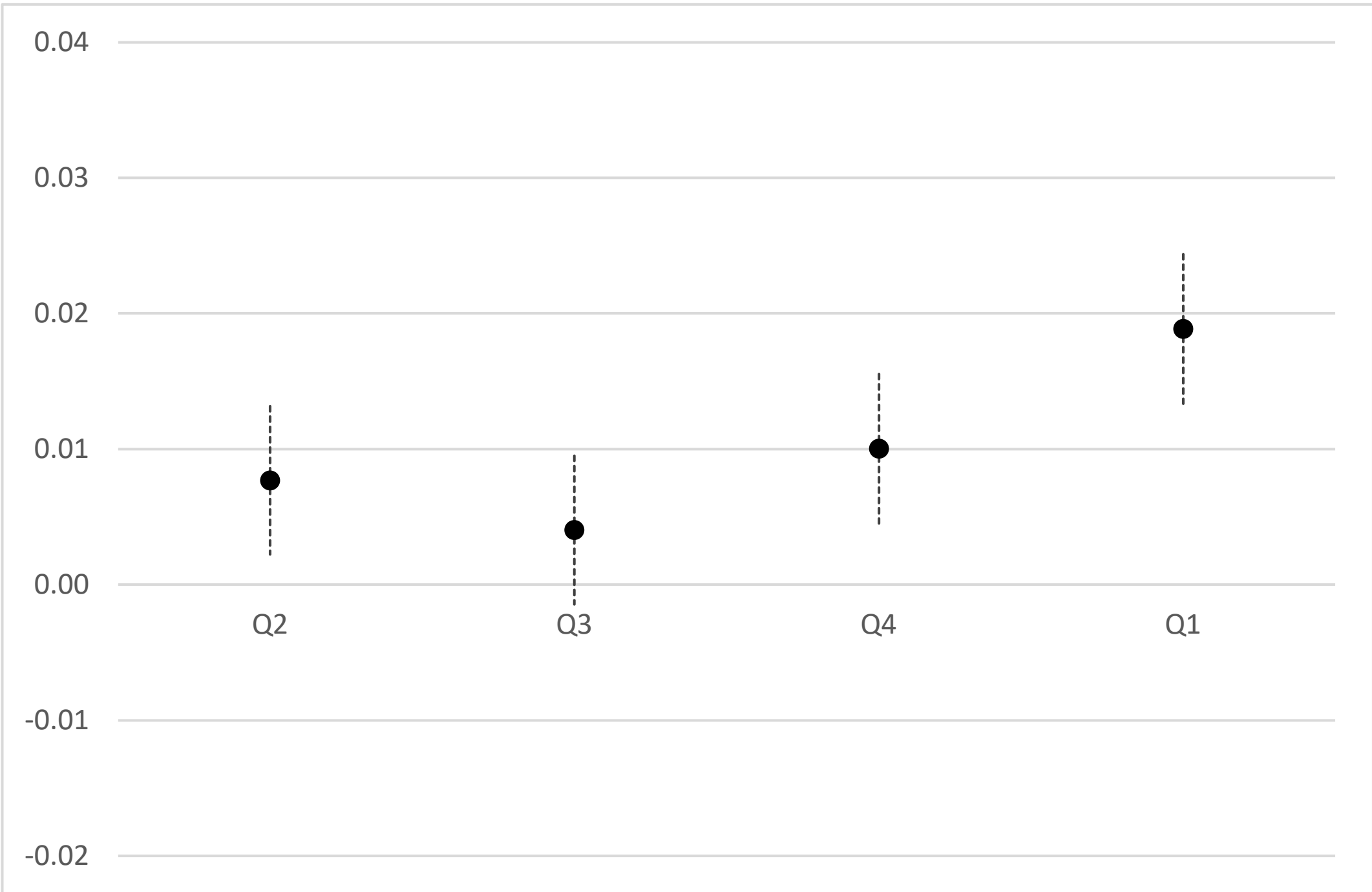

Panel B: Estimated effects of the reform

*Note:* Panel A shows the share of children with at least one report to child protictive services; Panel B shows the estimates and 95%-confidence intervals from a comparative event study estimation anchored in Q1, the quarter just prior to the passage of the reform. 2016 (2015) population consists of 36,859 (39,228) children of mothers on welfare benefits in

March 2016 (2015). The dotted vertical line indicates the passage of the reform (Q1 2016 (2015) for the 2016 (2015) population); the solid vertical line indicates the timing of the reform implementation (Q3 2016 (2015) for the 2016 (2015) population). The full set of estimates is shown in Table A12. Sample size: 577,256.

Figure 6

Estimated effects of the reform on mothers' outcomes, triple comparative event study

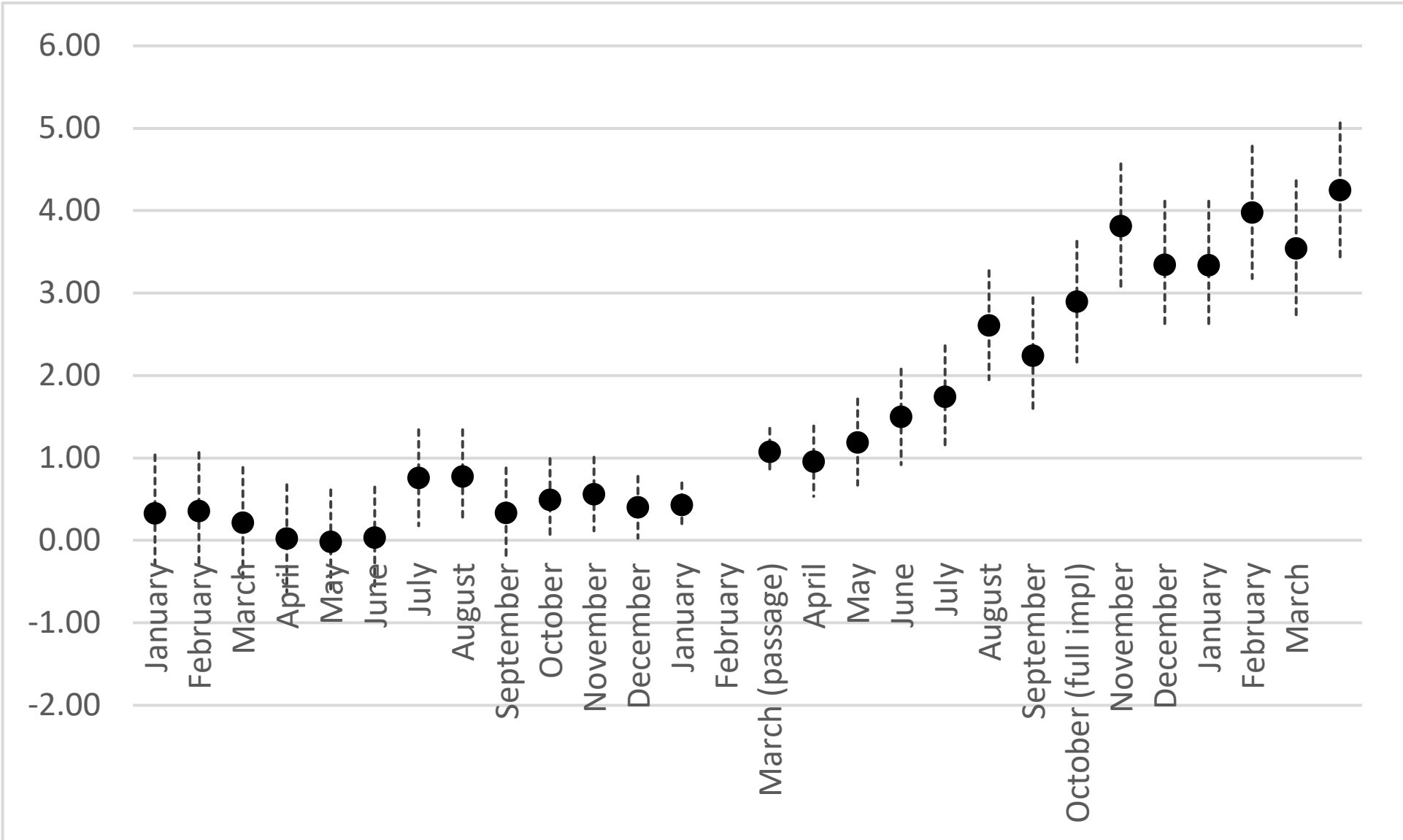

Panel A: Mothers' hours worked

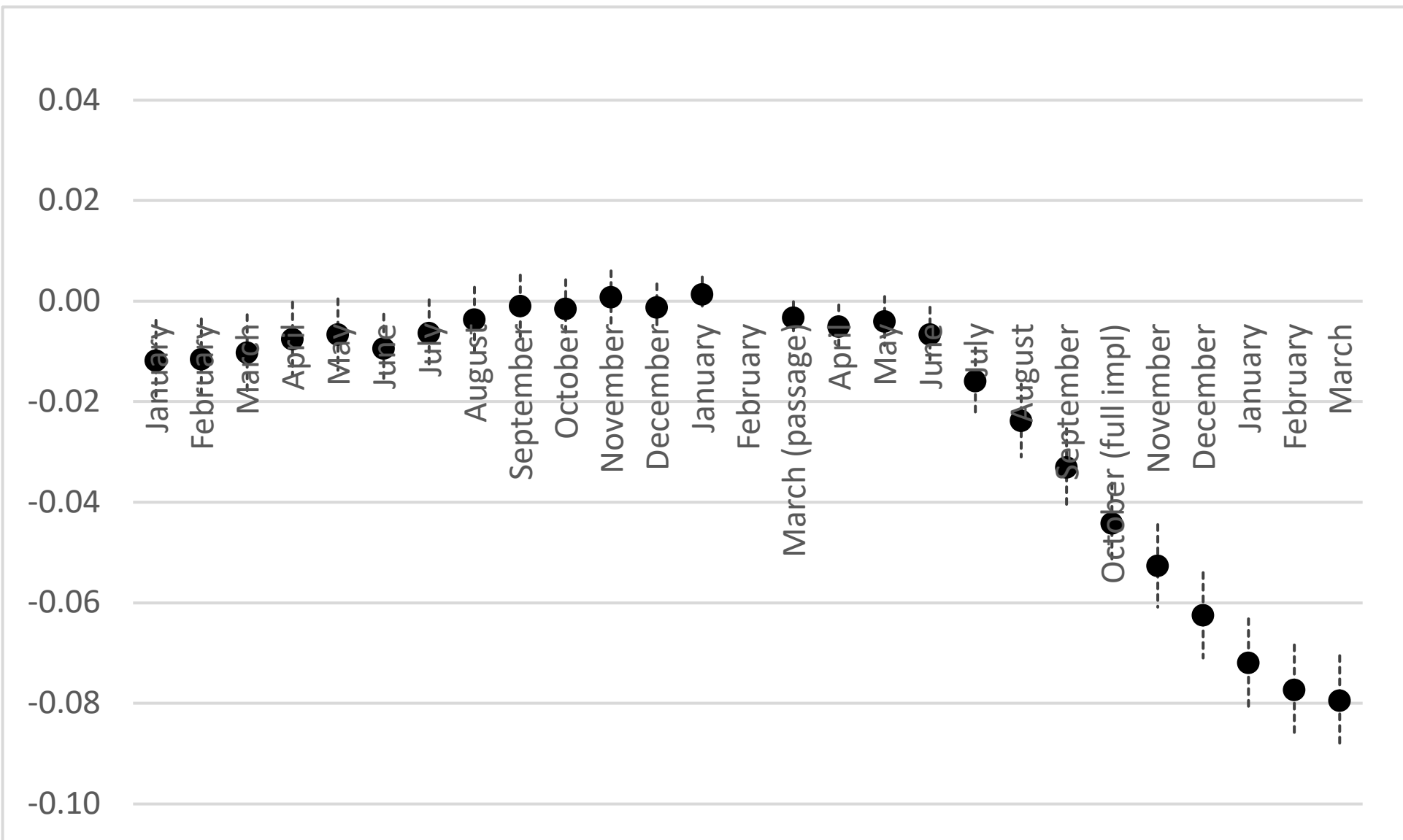

Panel B: Mothers' welfare participation

*Note:* This figure shows the estimates and 95%-confidence intervals from a triple comparative event study anchored in February, just prior to the reform. The approach adds the population of mothers not on welfare benefits in March 2015 and March 2016. Panel A shows the results for mothers' hours worked and Panel B those for welfare participation. The full set of estimates is shown in Table A6. 18,578 mothers on welfare make up the 2016 population exposed to the reform, while 19,716 mothers on welfare make up the 2015 comparison population. 616,310 mothers are in the non-welfare 2016 population and 616,789 mothers are in the non-welfare 2015 population.

Table 1

Descriptive statistics, at risk mothers on welfare

| | Welfare recipients 2015 | Welfare recipients 2016 | Standardized difference | Overall population, adult women 2016 | Standardized difference |
|---|---|---|---|---|---|
| Sample (really population) size | 19,716 | 18,578 | | 616,310 | |
| Age | 37.57 | 37.79 | 0.03 | 40.50 | -0.34 |
| | (8.3) | (8.5) | | (7.5) | |
| Number of children | 2.48 | 2.47 | -0.01 | 2.08 | 0.33 |
| | (1.4) | (1.4) | | (0.9) | |
| Number of children < 18 years of age | 2.01 | 2.00 | -0.01 | 1.76 | 0.25 |
| | (1.1) | (1.1) | | (0.8) | |
| Household size | 3.10 | 3.09 | -0.01 | 3.77 | -0.55 |
| | (1.4) | (1.4) | | (1.1) | |
| Dwelling size (square meter) | 102.63 | 102.56 | 0.00 | 136.05 | -0.57 |
| | (42.6) | (42.1) | | (71.9) | |
| Living arrangement (%) | | | | | |
| ... Married | 11 | 11 | -0.02 | 61 | -1.23 |
| ... Living w/ father of (at least one) child | 9 | 9 | 0.01 | 16 | -0.21 |
| ... Living w/ another adult | 2 | 2 | 0.00 | 4 | -0.11 |
| ...Single | 78 | 78 | 0.01 | 19 | 1.46 |
| Ethnicity (%) | | | | | |
| ... Danish | 68 | 67 | -0.02 | 85 | -0.44 |
| ... Immigrant | 29 | 30 | 0.02 | 13 | 0.41 |
| ... Immigrant descent | 3 | 3 | 0.02 | 1 | 0.12 |
| Educational attainment, % | | | | | |
| ... primary and lower secondary | 60 | 60 | 0.00 | 12 | 1.15 |
| … upper secondary | 29 | 29 | -0.01 | 38 | -0.20 |
| ... some tertiary | 6 | 7 | 0.02 | 49 | -1.07 |
| ... missing | 5 | 5 | 0.00 | 1 | 0.22 |
| Years of schooling | 11.39 | 11.41 | 0.01 | 14.77 | -1.32 |
| | (2.4) | (2.5) | | (2.6) | |
| Labor market experience, years | 3.39 | 3.28 | -0.02 | 12.74 | -1.36 |
| | (4.8) | (4.7) | | (8.6) | |
| Hours worked, March | | | | | |
| ... Total | 3.34 | 3.00 | -0.02 | 111.30 | -2.16 |
| | (18.7) | (16.9) | | (68.9) | |
| ... Unsubsidized | 2.77 | 2.63 | -0.01 | 109.61 | -2.10 |
| | (16.7) | (15.5) | | (70.5) | |
| Household income March, household equivalence (€) | | | | | |
| Labor income | 117 | 108 | -0.02 | 3736 | -0.85 |
| | (436) | (381) | | (5,989) | |
| Government transfers | 1156 | 1162 | 0.02 | 256 | 0.71 |
| | (292) | (436) | | (1,739) | |
| Discretionary income | 822 | 812 | -0.02 | N/A | |
| | (433) | (499) | | | |
| Assets and liabilities, household equivalence (€) | | | | | |
| owner of property (%) | 2.95 | 2.49 | -0.03 | 71.91 | -2.06 |
| ... Value of property | 2198 | 1953 | -0.02 | 136077 | -0.64 |
| | (16,568) | (15,156) | | (297,354) | |
| … value of mortgage | 1050 | 825 | -0.03 | 87078 | -0.67 |
| | (9,376) | (8,141) | | (181,272) | |
| Owner of vehicle(s) (%) | 26 | 23 | -0.06 | 79 | -1.34 |
| ... Market value of car | 886 | 360 | -0.29 | 4083 | -1.08 |
| | (2,232) | (1,193) | | (4,725) | |
| Financial assets (excl. pensions) (€) | 25 | 16 | -0.01 | 8120 | -0.04 |
| | (859) | (571) | | (270,343) | |
| Pensions (€) | 7245 | 6980 | -0.03 | 56769 | -1.07 |
| | (11,779) | (8,733) | | (65,209) | |
| Bank deposits (€) | 916 | 901 | -0.01 | 12238 | -0.43 |
| | (2,639) | (1,926) | | (36,848) | |
| Unsecured loans (€) | 10782 | 10851 | 0.00 | 25648 | -0.20 |
| | (21,525) | (19,752) | | (104,682) | |
| Medical utilization previous 12 months, any (%) | | | | | |
| Inpatient hospital care: | | | | | |
| ... Psychiatric | 2.0 | 1.8 | -0.01 | 0.3 | 0.15 |
| ... Related to reproduction | 11 | 11 | 0.00 | 9 | 0.06 |
| ...(other) Somatic | 16 | 15 | -0.02 | 8 | 0.24 |
| Outpatient hospital Care: | | | | | |
| ... Psychiatric | 11 | 11 | -0.01 | 2 | 0.39 |
| ... Related to reproduction | 15 | 15 | 0.00 | 11 | 0.11 |
| ...(other) Somatic | 53 | 53 | 0.00 | 40 | 0.27 |
| Primary care provider visit | 97 | 97 | -0.01 | 93 | 0.20 |
| Specialist visit | 41 | 41 | 0.01 | 35 | 0.13 |
| Urgent Care visit | 22 | 23 | 0.02 | 13 | 0.26 |
| Dentist visit | 36 | 37 | 0.01 | 63 | -0.54 |
| Crime previous 12 months, any (%) | | | | | |
| Victim of | | | | | |
| ... Sexual assault | 0.1 | 0.2 | 0.02 | < 0.1 | 0.05 |
| ... Violent assault | 1.6 | 1.6 | 0.00 | 0.3 | 0.14 |
| ... Property crime | 1.5 | 1.6 | 0.00 | 0.8 | 0.07 |
| charged with | | | | | |
| ... Sexual assault | < 0.1 | < 0.1 | 0.00 | < 0.1 | 0.02 |
| ... Violent assault | 0.5 | 0.6 | 0.00 | 0.1 | 0.09 |
| ... Property crime | 2.6 | 2.3 | -0.02 | 0.2 | 0.19 |
| ... DUI | 0.3 | 0.3 | -0.01 | < 0.1 | 0.06 |
| ... Possesion w/ intent to distribute | 0.4 | 0.4 | -0.01 | < 0.1 | 0.08 |

*Notes:* The table shows descriptive statistics for the reform and comparison groups and compares these to the overall population of women aged 18 or above who do not receive any welfare benefits in March 2016. Labor market experience is calculated using third-party reporting of hours worked from a mandatory supplementary pension scheme (ATP) introduced in April, 1964. Mortgage and property values include zeros. We define family-size-adjusted discretionary income as the sum of mother's and any partner's earnings and transfers minus taxes and rent, divided by the square root of family size; see OECD (2013) for a discussion of choice of equivalence scale.

Table 2

Descriptive statistics, children

| | Children of welfare recipients | | Standardized difference | Overall population, children | Standardized difference |
|---|---|---|---|---|---|
| | 2015 | 2016 | | 2016 | |
| Sample (really population) size | 39,228 | 36,859 | | 1,072,809 | |
| Age | 9.56 | 9.59 | 0.01 | 9.34 | 0.05 |
| | (5.0) | (5.0) | | (5.1) | |
| Female, % | 49 | 49 | -0.01 | 49 | 0.00 |
| Age group, % | | | | | |
| ... 0 - 5 years old | 27 | 27 | 0.00 | 30 | -0.06 |
| ... 6 - 12 years old | 43 | 43 | 0.00 | 41 | 0.04 |
| ... 13 - 17 years old | 30 | 30 | 0.01 | 29 | 0.02 |
| Ethnicity, % | | | | | |
| ... At least one parent is Danish citizen | 65 | 65 | -0.01 | 90 | -0.64 |
| ... Born abroad to non-Danish parents | 2 | 2 | 0.00 | 3 | -0.04 |
| ... Born in DK to non-Danish parents | 32 | 33 | 0.01 | 7 | 0.69 |
| Living arrangement, % | | | | | |
| ... Living w/ both parents | 22 | 22 | 0.00 | 75 | -1.26 |
| ... Living w/ mother and her new partner | 4 | 4 | -0.01 | 6 | -0.13 |
| ... Living w/ single mother | 62 | 63 | 0.01 | 15 | 1.12 |
| ... Living w/ father and his new partner | 2 | 2 | 0.00 | 1 | 0.11 |
| ... Living w/ single father | 5 | 4 | -0.01 | 2 | 0.13 |
| ... Living w/ neither parent | 6 | 6 | 0.01 | 1 | 0.29 |
| Information from medical birth records | | | | | |
| % with records | 96 | 96 | | 94 | |
| Mother's weight and height recorded, %* | 54 | 59 | | 60 | |
| Birth weight, grams | 3,352 | 3,346 | -0.01 | 3,487 | -0.23 |
| | (615) | (611) | | (610) | |
| Birth weight < 2,500 grams, % | 7 | 7 | 0.00 | 5 | 0.08 |
| Gestation age, days | 276 | 276 | 0.00 | 278 | -0.12 |
| | (14) | (14) | | (14) | |
| Gestation age < 224 days (32 weeks), % | 1.19 | 1.13 | -0.01 | 0.90 | 0.02 |
| APGAR score | 9.86 | 9.86 | 0.00 | 9.87 | -0.01 |
| | 0.66 | 0.65 | | 0.61 | |
| Number of prenatal visits to midwife | 4.70 | 4.58 | -0.05 | 4.93 | -0.16 |
| | (2.2) | (2.3) | | (2.1) | |
| Mother's BMI prior to pregnancy | 25.51 | 25.49 | 0.00 | 24.40 | 0.13 |
| | (9.1) | (9.2) | | (7.9) | |
| Mother's BMI > 30, % | 21 | 20 | 0.00 | 12 | 0.23 |
| Mother smoking during pregnancy, % | 43 | 43 | -0.01 | 17 | 0.59 |
| Father's age at time of birth | 31.01 | 31.11 | 0.01 | 32.75 | -0.25 |
| | (7.2) | (7.3) | | (5.7) | |
| Father teenager at time of birth, % | 2 | 2 | 0.00 | 0.31 | 0.18 |
| Mother's age at time of birth | 27.42 | 27.59 | 0.03 | 30.23 | -0.49 |
| | (5.9) | (5.9) | | (4.8) | |
| Mother teenager at time of birth, % | 7 | 7 | -0.01 | 1 | 0.30 |
| Number of children of school age | 25,016 | 23,360 | | 635,890 | |
| School form, % | | | | | |
| ... Public School | 80 | 79 | -0.01 | 78 | 0.04 |
| ... Private school | 10 | 10 | 0.02 | 18 | -0.22 |
| ... Special ed school | 5 | 5 | 0.01 | 2 | 0.19 |
| ... Home schooled | 6 | 5 | -0.02 | 3 | 0.12 |
| Old for grade, % ** | 27 | 25 | -0.04 | 14 | 0.30 |
| New school current school year, %** | 9 | 11 | 0.06 | 6 | 0.18 |
| Test scores taken in grade 2, 4, 6, and 8 | | | | | |
| Reading | | | | | |
| ...N in sampling year | 7,012 | 7,510 | | 192,403 | |
| ... Standardized score | -0.53 | -0.57 | 0.03 | 0.06 | -0.43 |
| | (0.97) | (0.97) | | (0.96) | |
| Math | | | | | |
| ...N in sampling year | 3,709 | 3,879 | | 99,273 | |
| … Standardized score | -0.60 | -0.61 | 0.01 | 0.06 | -0.48 |
| | (0.97) | (0.97) | | (0.97) | |
| Well-being, grades 4-9 | | | | | |
| ...N in sampling year | 9,524 | 8,872 | | 264,839 | |
| Well-being average (1-5) | 4.00 | 3.94 | 0.06 | 4.14 | -0.15 |
| | (0.68) | (0.68) | | (0.61) | |
| Reports to child protective services | | | | 1,068,477 | |
| … N in sampling year | 36,859 | 39,228 | | | |
| Any report in Q1 | 0.073 | 0.080 | -0.02 | 0.012 | 0.22 |

*Note:* The table shows descriptive statistics for children of mothers on welfare in March 2015 and March 2016 and included in our analysis, along with the overall population of Danish children. Note that in Denmark, unlike the US, essentially all mothers make use of a midwife. Test scores, self-reported social well-being, and reports to child protective services are measured in the spring of 2015 (2016) for the 2015 (2016) population. The APGAR test is carried out by a doctor, midwife, or nurse shortly after the birth and rates the baby's breathing effort, heart rate, muscle tone, reflexes, and skin color. Each category is scored with 0, 1, or 2, depending on the observed condition.

* Mothers' height and weight are only recorded starting in 2004.

** Conditional on private or public school attendance.

Table 3

Estimated effects of the reform on test scores

| | Danish reading | | Math | |
|---|---|---|---|---|
| Variables | Coef. | Std. error | Coef. | Std. error |
| Reform population indicator | 0.031 | 0.018 | 0.019 | 0.022 |
| Time indicators: | | | | |
| One year prior | 0.000 | 0.017 | 0.034 | 0.022 |
| One year after | **0.034** | 0.017 | -0.021 | 0.022 |
| Effects of the reform: | | | | |
| One year prior | -0.023 | 0.031 | -0.056 | 0.037 |
| One year after | 0.015 | 0.030 | 0.031 | 0.037 |
| Constant | **-0.569** | 0.012 | **-0.614** | 0.016 |
| # children, reform pop. | 17,240 | | 11,422 | |
| # children, comparison pop. | 18,136 | | 12,249 | |

*Notes:* The table shows the results from comparative event study estimation using 2015-2017 data for the reform cohort and 2014-2016 data for the comparison cohort. The reform (comparison) cohort consists of children of mothers on welfare in March 2016 (2015), enrolled in public schools and in tested grade levels. The analysis is anchored in Q1, just prior to the passage of the reform. Post-measurement is Q1 2017 (2016 for the comparison cohort). **Bold** indicates significance at a 5% level; *italic* indicates significance at a 10% level.

Table 4

Estimated effects of the reform on self-reported social well-being

| | (1) | | (2) | |
|---|---|---|---|---|
| Variables | Coef. | Std. error | Coef. | Std. Error |
| Reform population indicator | 0.035 | 0.006 | 0.009 | 0.005 |
| Time indicator: | | | | |
| One year after | 0.059 | 0.008 | 0.010 | 0.005 |
| Effect of the reform: | | | | |
| One year after | **-0.032** | 0.010 | **-0.055** | 0.013 |
| Time trend | | | **0.050** | 0.008 |
| Constant | **3.94** | 0.007 | **3.99** | 0.008 |
| # children, reform pop. | | 14,729 | | |
| # children, comparison pop. | | 13,170 | | |

*Notes:* The table shows the results from comparative event study estimation using 2015-2017 data for reform cohort and 2015-2016 data for comparison cohort. The reform (comparison) cohort consists of children of mothers on welfare in March 2016 (2015) who were enrolled in public schools. The analysis is anchored in Q1, just prior to the passage of the reform. Post-measurement is Q1 2017 (2016) for the reform (comparison) cohort. **Bold** indicates significance at a 5% level; *italic* indicates significance at a 10% level.

Table 5

Estimated effects of the reform on childrens' outcomes, triple comparative event study

| | Coef. | Std. error |
|---|---|---|
| Panel A. Danish reading | | |
| Reform population indicator | 0.0002 | 0.0031 |
| Welfare population indicator | **-0.629** | 0.012 |
| Reform X welfare population indicator | 0.031 | 0.018 |
| | | |
| Times dummies: | | |
| One year prior | -0.004 | 0.003 |
| One year after | 0.001 | 0.003 |
| | | |
| Time dummies X reform population indicator: | | |
| One year prior | 0.003 | 0.006 |
| One year after | -0.001 | 0.006 |
| | | |
| Time dummies X welfare population indicator: | | |
| One year prior | 0.004 | 0.018 |
| One year after | *0.033* | 0.017 |
| | | |
| Effects of the reform: | | |
| One year prior | -0.026 | 0.031 |
| One year after | 0.015 | 0.031 |
| | | |
| # children, reform population | 17,240 | |
| # children, comparison population | 18,136 | |
| # non-welfare 2016 children | 453,159 | |
| # non-welfare 2015 children | 454,386 | |
| | | |
| Panel B. Math | | |
| Reform population indicator | -0.001 | 0.004 |
| Welfare population indicator | **-0.672** | 0.016 |
| Reform X welfare population indicator | 0.020 | 0.023 |
| | | |
| Times dummies: | | |
| One year prior | -0.004 | 0.004 |
| One year after | 0.001 | 0.004 |
| | | |
| Time dummies X reform population indicator: | | |
| One year prior | 0.004 | 0.008 |
| One year after | 0.000 | 0.008 |
| | | |
| Time dummies X welfare population indicator: | | |
| One year prior | 0.038 | 0.022 |
| One year after | -0.022 | 0.022 |
| Effects of the reform: | | |
| One year prior | -0.059 | 0.038 |
| One year after | 0.030 | 0.038 |
| | | |
| # children, reform population | 11,422 | |
| # children, comparison population | 12,249 | |
| # non-welfare 2016 children | 301,225 | |
| # non-welfare 2015 children | 301,523 | |
| | | |
| Panel C. Social well-being | | |
| Reform population indicator | -0.0003 | 0.0002 |
| Welfare population indicator | **-0.142** | 0.007 |
| Reform X welfare population indicator | 0.004 | 0.006 |
| | | |
| Times dummies: | | |
| One year after | -0.0001 | 0.0003 |
| | | |
| Time dummies X reform population indicator: | | |
| One year after | **-0.059** | 0.002 |
| | | |
| Time dummies X welfare population indicator: | | |
| One year after | 0.001 | 0.008 |
| | | |
| Effects of the reform: | | |
| One year after | 0.0001 | 0.0103 |
| | | |
| # children, reform population | 14,729 | |
| # children, comparison population | 13,170 | |
| # non-welfare 2016 children | 388,754 | |
| # non-welfare 2015 children | 330,797 | |
| | | |
| Panel D. Any report to child protective services | | |
| Reform population indicator | **0.0006** | **0.0001** |
| Welfare population indicator | **0.068** | 0.001 |
| Treated X welfare population indicator | *-0.003* | 0.002 |
| | | |
| Time dummies: | | |
| 1st quarter after | **-0.0004** | 0.0001 |
| 2nd quarter after | **-0.0017** | 0.0001 |
| 3rd quarter after | **0.0006** | 0.0001 |
| 4th quarter after | **0.0010** | 0.0001 |
| | | |
| Time dummies X reform population indicator: | | |
| 1st quarter after | **0.0018** | 0.0002 |
| 2nd quarter after | 0.0002 | 0.0002 |
| 3rd quarter after | **0.0010** | 0.0002 |
| 4th quarter after | **0.0020** | 0.0003 |
| Time dummies X welfare population indicator: | | |
| 1st quarter after | -0.001 | 0.002 |
| 2nd quarter after | **-0.009** | 0.002 |
| 3rd quarter after | **-0.004** | 0.002 |
| 4th quarter after | **-0.007** | 0.002 |
| | | |
| Effects of the reform | | |
| 1st quarter after | **0.006** | 0.003 |
| 2nd quarter after | 0.004 | 0.003 |
| 3rd quarter after | **0.009** | 0.003 |
| 4th quarter after | **0.017** | 0.003 |
| | | |
| # children, reform population | 36,859 | |
| # children, comparison population | 39,228 | |
| # non-welfare 2016 children | 1,073,923 | |
| # non-welfare 2015 children | 1,069,701 | |

*Note:* This table shows the estimates and 95%-confidence intervals from a triple comparative event study that adds population of children of mothers not on welfare benefits in March 2015 and March 2016. The analyses are all anchored in the quarter just prior to the passage of the reform. Post-measurements for Danish reading and social well-being are taken in Q1 2017 (2016 for the comparison cohort). Post-measurements for child protective services are taken in the four quarters following the passage of the reform. The social well-being model includes a linear time trend but this does not impact the estimated effect of the reform. **Bold** indicates significance at a 5% level; *italic* indicates significance at a 10% level.

Table 6

Heterogeneity in the estimated effects of the reform, by the mothers' relationship status

| | Mother single | | Mother married or cohabiting | | P-values equal subgroup |
|---|---|---|---|---|---|
| | Coef. | Std. error | Coef. | Std. error | effects |
| **Child level analyses** | | | | | |
| Danish reading | | | | | |
| One year after | -0.005 | 0.035 | 0.072 | 0.058 | 0.2527 |
| Math | | | | | |
| One year after | 0.004 | 0.043 | *0.117* | 0.070 | 0.1671 |
| Social well-being | | | | | |
| One year after | **-0.051** | 0.015 | **-0.067** | 0.024 | 0.5684 |
| Child protective services | | | | | |
| 1st quarter after | **0.010** | 0.003 | 0.001 | 0.004 | 0.0750 |
| 2nd quarter after | *0.005* | 0.003 | 0.001 | 0.004 | 0.3939 |
| 3rd quarter after | **0.012** | 0.003 | 0.005 | 0.004 | 0.1729 |
| 4th quarter after | **0.019** | 0.004 | **0.019** | 0.005 | 0.9648 |
| **Family level analyses** | | | | | |
| Family adjusted discretionary income (€) | | | | | |
| One year after | **-323.36** | 5.66 | **-289.04** | 15.33 | |

*Notes:* The table shows heterogeneity by mothers' relationship status. The specifications parallel those from Tables 3-4 and Figure 5. Bold indicates significance at a 5% level; italic indicates significance at a 10% level.

Table 7

Heterogeneity in the estimated effects of the reform, by the children's sex and age

| | Girls | | Boys | | P-values equal subgroup | Age < 6 | | Age 7-14 | | Age 15-18 | | P-values equal subgroup |
|---|---|---|---|---|---|---|---|---|---|---|---|---|
| | Coef. | Std. error | Coef. | Std. error | effects | Coef. | Std. error | Coef. | Std. error | Coef. | Std. error | effects |
| **<u>Child level analyses</u>** | | | | | | | | | | | | |
| Danish reading | | | | | | | | | | | | |
| One year after | 0.037 | 0.041 | -0.012 | 0.045 | 0.4184 | | | | | | | |
| Math | | | | | | | | | | | | |
| One year after | 0.023 | 0.051 | 0.037 | 0.054 | 0.8519 | | | | | | | |
| Social well-being | | | | | | | | | | | | |
| One year after | **-0.050** | 0.018 | **-0.060** | 0.017 | 0.7076 | | | | | | | |
| Child protective services | | | | | | | | | | | | |
| 1st quarter after | 0.004 | 0.004 | **0.011** | 0.004 | 0.1653 | 0.000 | 0.005 | **0.013** | 0.003 | 0.003 | 0.006 | 0.0739 |
| 2nd quarter after | 0.001 | 0.004 | **0.006** | 0.004 | 0.3116 | -0.001 | 0.005 | **0.007** | 0.003 | 0.003 | 0.005 | 0.4184 |
| 3rd quarter after | 0.008 | 0.004 | **0.012** | 0.004 | 0.4298 | 0.006 | 0.005 | **0.013** | 0.004 | 0.005 | 0.006 | 0.3181 |
| 4th quarter after | 0.009 | 0.004 | **0.028** | 0.004 | 0.0030 | **0.012** | 0.006 | **0.021** | 0.004 | **0.020** | 0.007 | 0.4705 |
| **<u>Family level analyses</u>** | | | | | | | | | | | | |
| Family adjusted discretionary income (€) | | | | | | | | | | | | |
| One year after | **-316.94** | 7.59 | **-316.99** | 7.14 | | **-277.53** | 9.45 | **-330.28** | 6.90 | **-355.59** | 10.25 | |

*Notes:* The table shows heterogeneity by children's sex and age. The specifications parallel those from Tables 3-4 and Figure 5. Subgroup overlap occurs in the family level analyses. Bold indicates significance at a 5% level; italic indicates significance at a 10% level.

## Appendix A

Figure A1

Estimated effects of the reform by month, partners' outcomes

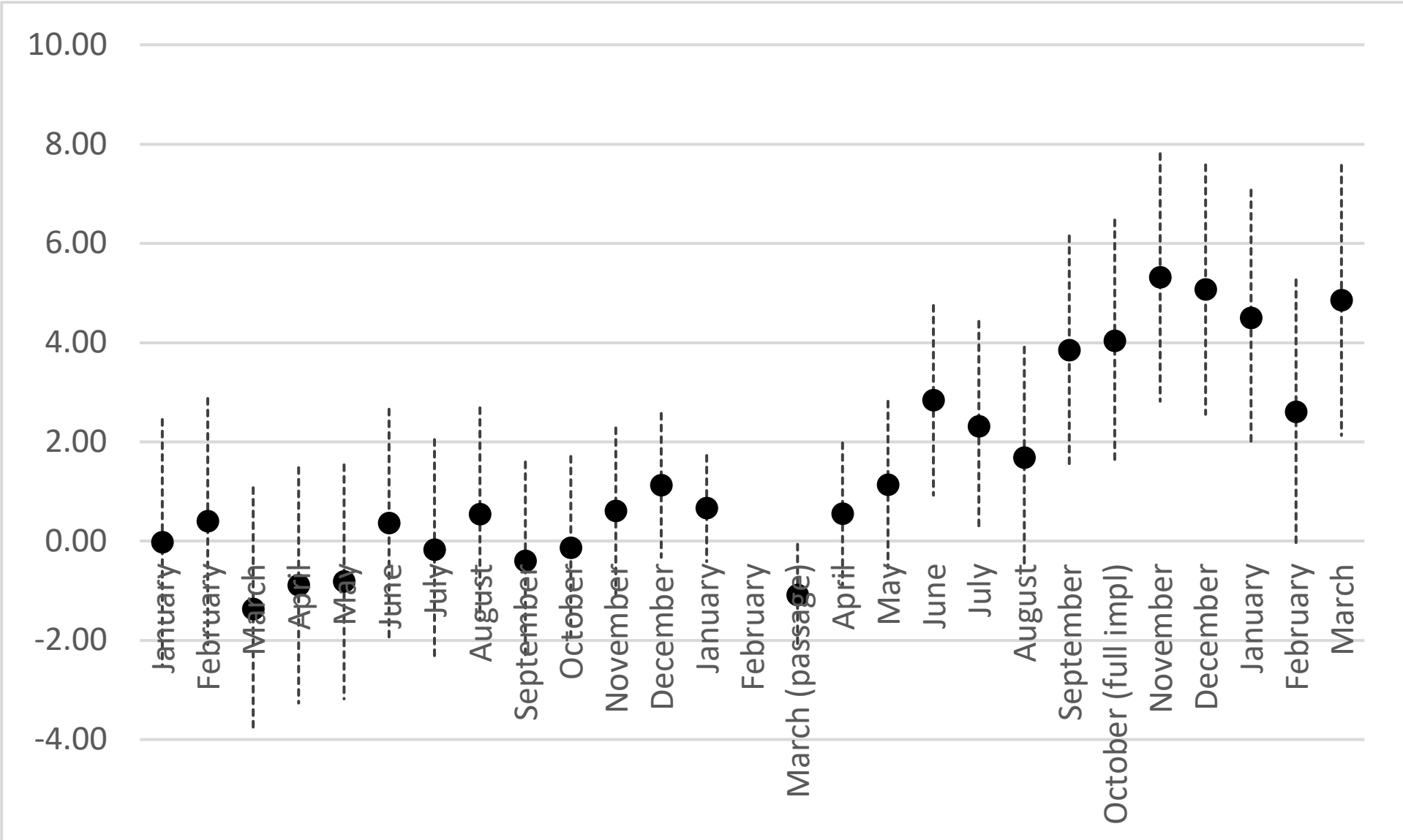

Panel A: Hours worked

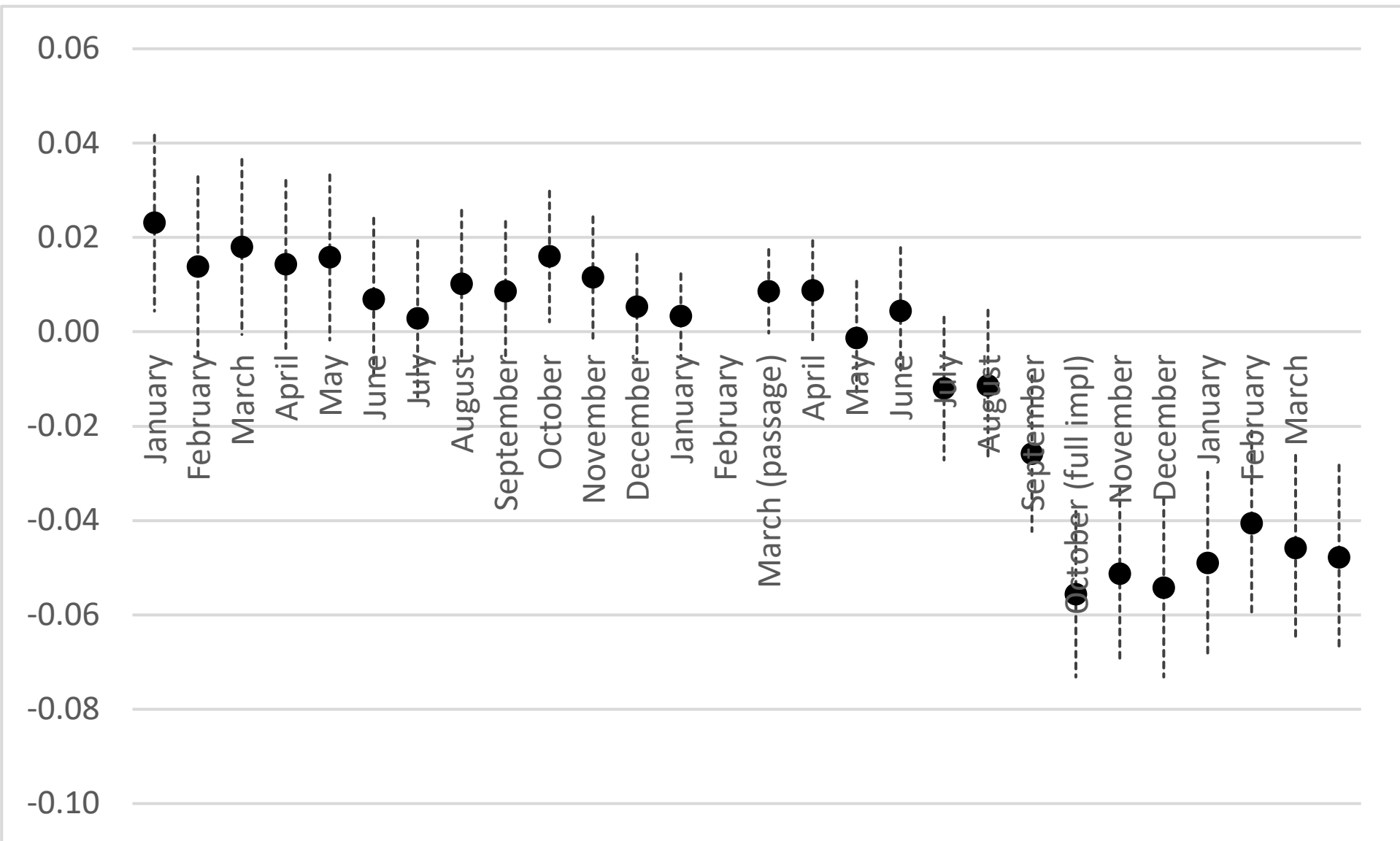

Panel B: Welfare

*Notes:* This figure shows the estimates and 95%-confidence intervals from a comparative event study estimation anchored in February, just prior to the reform. 2016 (2015) population consists of partners to mothers on welfare benefits in March 2016 (2015). Sample size: 303,264.

Figure A2

Inflow into welfare

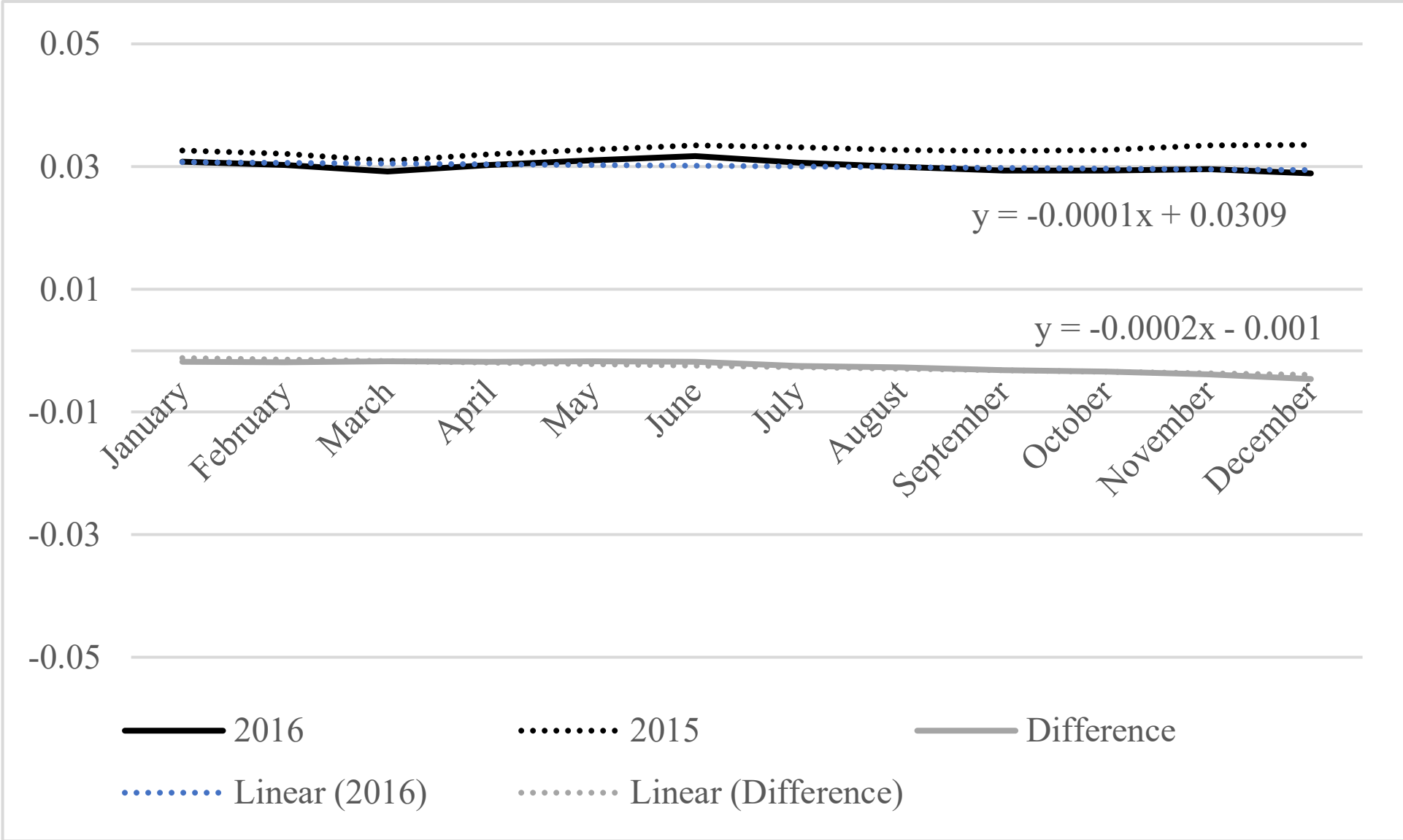

*Note:* The figure shows the share of mothers on welfare benefits during 2015 and 2016 as well as the difference between the shares. Estimated trend lines relate to the share of welfare in 2016 (upper) and the difference between the 2016 and 2015 shares.

Table A1

Predicted pre- and post-reform disposable income absent behavioral changes, by family types

| | Pre-reform disposable income | Post-reform disposable income | Percentage change |
|---|---|---|---|
| *Singles* | | | |
| No children | 5,300 | 5,300 | 0% |
| One child | 9,700 | 7,400 | -24% |
| Two children | 13,100 | 10,500 | -20% |
| Three children | 16,500 | 13,100 | -21% |
| Four children | 20,500 | 16,200 | -21% |
| *Cohabiting or married couples* | | | |
| No children | 10,700 | 9,200 | -14% |
| One child | 14,500 | 13,800 | -5% |
| Two children | 15,600 | 14,400 | -8% |
| Three children | 17,600 | 15,300 | -13% |
| Four children | 19,500 | 15,600 | -20% |

*Notes:* This table shows disposable monthly income in DKK after housing costs for individuals aged 30 or older. Calculation assumes monthly rent for single without children of DKK 2,801 and DKK 6,138 for other family types. Calculation assumes other costs of housing to amount to DKK 761 and DKK 1,296. In families with one child, the child is assumed to be five years old; in families with two children five and ten years old; in families with three children five, ten, and 14 years old; and in families with four children one, five, ten, and 14 years old. Source: The Danish Ministry of Employment (2015b).

Table A2

Comparing mothers at risk of facing the upper limit on benefits and other mothers on welfare

| | 2015 | | 2016 | |
|---|---|---|---|---|
| | At risk of facing upper limit on total benefits | Not at risk | At risk of facing upper limit on total benefits | Not at risk |
| Assets and liabilities, household equivalence (€) | | | | |
| owner of property (%) | 3.0 | 23.5 | 2.5 | 22.9 |
| ... Value of property | 2,198 | 23,266 | 1,953 | 21,355 |
| | (16,568) | (247,122) | (15,156) | (71,851) |
| … value of mortgage | 1,050 | 13,664 | 825 | 12,384 |
| | (9,376) | (98,461) | (8,141) | (56,542) |
| Owner of vehicle(s) (%) | 26 | 49 | 23 | 47 |
| ... Market value of car | 886 | 2,362 | 360 | 1,091 |
| | (2,232) | (3.958) | (1,193) | (2,270) |
| Financial assets (excl. pensions) (€) | 25 | 266 | 16 | 366 |
| | (859) | (11,509) | (571) | (13,202) |
| Pensions (€) | 7,245 | 13,799 | 6,980 | 14,162 |
| | (11,779) | (16,470) | (8,733) | (17,568) |
| Bank deposits (€) | 916 | 1,678 | 901 | 1,878 |
| | (2,639) | (5,647) | (1,926) | (7,829) |
| Unsecured loans (€) | 10,782 | 14,297 | 10,851 | 14,719 |
| | (21,525) | (36,325) | (19,752) | (31,734) |
| Medical utilization previous 12 months, any (%) | | | | |
| Inpatient hospital care: | | | | |
| ... Psychiatric | 2.0 | 2.2 | 1.8 | 1.7 |
| ... Related to reproduction | 11 | 18 | 11 | 17 |
| ...(other) Somatic | 16 | 18 | 15 | 16 |
| Outpatient hospital Care: | | | | |
| ... Psychiatric | 11 | 10 | 11 | 10 |
| ... Related to reproduction | 15 | 19 | 15 | 20 |
| ...(other) Somatic | 53 | 55 | 53 | 54 |
| Primary care provider visit | 97 | 97 | 97 | 96 |
| Specialist visit | 41 | 35 | 41 | 35 |
| Urgent Care visit | 22 | 24 | 23 | 26 |
| Dentist visit | 36 | 33 | 37 | 34 |
| Crime previous 12 months, any (%) | | | | |
| Victim of | | | | |
| ... Sexual assault | 0.1 | 0.1 | 0.2 | 0.1 |
| ... Violent assault | 1.6 | 1.5 | 1.6 | 1.3 |
| ... Property crime | 1.5 | 1.0 | 1.6 | 1.1 |
| charged with | | | | |
| ... Sexual assault | < 0.1 | < 0.1 | < 0.1 | < 0.1 |
| ... Violent assault | 0.5 | 0.5 | 0.6 | 0.1 |
| ... Property crime | 2.6 | 2.9 | 2.3 | 2.6 |
| ... DUI | 0.3 | 0.4 | 0.3 | 0.4 |
| ... Possesion w/ intent to distribute | 0.4 | 0.5 | 0.4 | 0.4 |

*Notes:* The table shows descriptive statistics for the reform and comparison groups (the groups at risk of facing the upper limit on benefits) and compares these to the population of mothers on welfare in March 2016 who do not receive some housing support. Labor market experience is calculated using third-party reporting of hours worked from a mandatory supplementary pension scheme (ATP) introduced in April, 1964. We define family-size-adjusted discretionary income as the sum of mother's and any partner's earnings and transfers minus taxes and rent, divided by the square root of family size; see OECD (2013) for a discussion of choice of equivalence scale.

Table A3

Sample loss journey

| | Individuals | | Percent | |
|---|---|---|---|---|
| | 2015 | 2016 | 2015 | 2016 |
| Total number of welfare recipients | 140,019 | 139,187 | 100% | 100% |
| With no other types of income support | 137,166 | 134,876 | 98% | 97% |
| Living in the country at the end of the previous year | 137,050 | 134,835 | 98% | 97% |
| Women | 65,535 | 64,771 | 47% | 47% |
| ... With children | 44,370 | 43,667 | 32% | 31% |
| … … Below the age of 18 | 34,827 | 33,960 | 25% | 24% |
| Living in country at least 7 out of 8 previous years | 31,098 | 30,175 | 22% | 22% |
| With benefit levels at risk of cuts | 19,716 | 18,578 | 14% | 13% |
| Number of children | 48,846 | 45,882 | | |
| … below 18 | 39,388 | 36,820 | | |

*Note:* The table shows how the selection criteria affect the number of observations in the sample.

Table A4

Descriptive statistics, fathers

| | Welfare recipients 2015 | Welfare recipients 2016 | Standardized difference | Overall population, fathers 2016 | Standardized difference |
|---|---|---|---|---|---|
| Sample size | 22,427 | 20,979 | | 603,338 | |
| Age | 40.54 | 40.68 | 0.02 | 43.03 | -0.27 |
| | (9.0) | (9.2) | | (8.0) | |
| Ethnicity (%) | | | | | |
| ... Danish | 71 | 70 | -0.02 | 87 | -0.43 |
| ... Immigrant | 27 | 27 | 0.02 | 12 | 0.40 |
| ... Immigrant descent | 2 | 3 | 0.02 | 1 | 0.12 |
| Educational attainment, % | | | | | |
| ... primary and lower secondary | 46 | 46 | 0.02 | 16 | 0.71 |
| … upper secondary | 36 | 35 | -0.02 | 45 | -0.21 |
| ... some tertiary | 8 | 8 | 0.01 | 37 | -0.72 |
| ... missing | 11 | 11 | 0.01 | 3 | 0.30 |
| Years of schooling | 12.2 | 12.2 | -0.01 | 14.5 | -0.91 |
| | (2.5) | (2.6) | | (2.6) | |
| Labor market experience, years | 9.4 | 9.2 | -0.02 | 16.6 | -0.80 |
| | (8.7) | (8.7) | | (9.7) | |
| Hours worked, March | | | | | |
| ... Total | 63.3 | 61.0 | -0.03 | 124.4 | -0.86 |
| | (78.3) | (78.3) | | (69.5) | |
| ... Unsubsidized | 62.8 | 60.8 | -0.02 | 123.4 | -0.85 |
| | (77.2) | (77.3) | | (70.6) | |
| Medical utilization previous 12 months, any (%) | | | | | |
| Inpatient hospital care: | | | | | |
| ... Psychiatric | 2 | 2 | 0.01 | 0.3 | 0.00 |
| ...(other) Somatic | 11 | 11 | 0.00 | 6 | 0.16 |
| Outpatient hospital Care: | | | | | |
| ... Psychiatric | 5 | 5 | 0.01 | 1 | 0.22 |
| ...(other) Somatic | 39 | 39 | 0.01 | 30 | 0.19 |
| Primary care provider visit | 77 | 77 | 0.00 | 76 | 0.04 |
| Specialist visit | 22 | 22 | 0.00 | 21 | 0.03 |
| Urgent Care visit | 15 | 16 | 0.03 | 10 | 0.18 |
| Dentist visit | 30 | 30 | 0.01 | 55 | -0.53 |
| Crime previous 12 months, any (%) | | | | | |
| Victim of | | | | | |
| ... Violent assault | 1.1 | 1.1 | 0.00 | 0.3 | 0.09 |
| ... Property crime | 0.7 | 0.8 | 0.01 | 0.5 | 0.04 |
| charged with | | | | | |
| ... Sexual assault | 0.2 | 0.2 | 0.01 | < 0.1% | . |
| ... Violent assault | 2.5 | 2.5 | 0.00 | 0.3 | 0.18 |
| ... Property crime | 4.9 | 4.8 | -0.01 | 0.6 | 0.26 |
| ... DUI | 2.2 | 2.1 | -0.01 | 0.3 | 0.16 |
| ... Possesion w/ intent to distribute | 3.3 | 2.6 | -0.04 | 0.3 | 0.20 |

*Notes:* The table shows descriptive statistics for the fathers of the children whose mothers are on welfare and compares these to the overall population of fathers aged 18 or above.

Table A5

Descriptive statistics, partners

| | Welfare recipients 2015 | Welfare recipients 2016 | Standardized difference | Overall population, partners 2016 | Standardized difference |
|---|---|---|---|---|---|
| Sample size | 4,362 | 4,063 | | 498,578 | |
| Age | 38.2 | 38.0 | -0.02 | 42.7 | -0.53 |
| | (9.6) | (9.7) | | (8.0) | |
| Male, % | 99.9 | 99.8 | 0.00 | 99.7 | 0.03 |
| Ethnicity (%) | | | | | |
| ... Danish | 52 | 51 | -0.01 | 87 | -0.83 |
| ... Immigrant | 44 | 45 | 0.01 | 12 | 0.77 |
| ... Immigrant descent | 4 | 4.06 | 0.01 | 1 | 0.19 |
| Educational attainment, % | | | | | |
| ... primary and lower secondary | 55 | 58 | 0.05 | 14 | 1.04 |
| … upper secondary | 31 | 28 | -0.06 | 45 | -0.36 |
| ... some tertiary | 8 | 8 | 0.01 | 40 | -0.81 |
| ... missing | 6.1 | 5.9 | -0.01 | 1.2 | 0.26 |
| Years of schooling | 11.7 | 11.6 | -0.03 | 14.7 | -1.15 |
| | (2.7) | (2.7) | | (2.6) | |
| Labor market experience, years | 5.4 | 4.9 | -0.08 | 16.6 | -1.48 |
| | (6.0) | (5.7) | | (9.6) | |
| Hours worked, March | | | | | |
| ... Total | 35.6 | 33.9 | -0.03 | 131.3 | -1.54 |
| | (62.8) | (61.4) | | (64.9) | |
| ... Unsubsidized | 32.2 | 31.7 | -0.01 | 130.4 | -1.56 |
| | (61.1) | (60.4) | | (66.0) | |
| Medical utilization previous 12 months, any (%) | | | | | |
| Inpatient hospital care: | | | | | |
| ... Psychiatric | 0.8 | 0.7 | -0.01 | 0.1 | 0.09 |
| ...(other) Somatic | 11 | 10 | -0.02 | 6 | 0.17 |
| Outpatient hospital Care: | | | | | |
| ... Psychiatric | 5 | 5 | 0.00 | 0.7 | 0.28 |
| ...(other) Somatic | 44 | 43 | -0.01 | 30 | 0.28 |
| Primary care provider visit | 88 | 86 | -0.04 | 76 | 0.26 |
| Specialist visit | 30 | 29 | 0.00 | 21 | 0.20 |
| Urgent Care visit | 18 | 20 | 0.05 | 10 | 0.28 |
| Dentist visit | 27 | 28 | 0.03 | 58 | -0.64 |
| Crime previous 12 months, any (%) | | | | | |
| Victim of | | | | | |
| ... Violent assault | 0.7 | 0.8 | 0.01 | 0.3 | 0.07 |
| ... Property crime | 0.4 | 0.6 | 0.03 | 0.4 | 0.03 |
| charged with | 0 | 0 | | 0 | |
| ... Sexual assault | <0.1% | 0.3 | 0.06 | < 0.1% | 0.07 |
| ... Violent assault | 2 | 2 | 0.01 | 0.2 | 0.17 |
| ... Property crime | 4 | 4 | 0.01 | 0.3 | 0.25 |
| ... DUI | 2 | 2 | 0.00 | 0.2 | 0.16 |
| ... Possesion w/ intent to distribute | 2 | 2 | -0.01 | 0.1 | 0.18 |

*Notes:* The table shows descriptive statistics for the partners to mothers who are on welfare and compares these to the overall population of male partners aged 18 or above.

Table A6

Estimated effects of the reform by month, mothers' outcomes

| | Outcome: hours worked | | | | Outcome: on welfare | | | |
|---|---|---|---|---|---|---|---|---|
| | Comparative event study | | Triple comparative event study | | Comparative event study | | Triple comparative event study | |
| | Coefficient | Standard error | Coefficient | Standard error | Coefficient | Standard error | Coefficient | Standard error |
| January | -0.061 | 0.344 | 0.332 | 0.359 | **-0.014** | 0.004 | **-0.012** | 0.004 |
| February | -0.127 | 0.344 | 0.358 | 0.359 | **-0.013** | 0.004 | **-0.012** | 0.004 |
| March | -0.239 | 0.327 | 0.215 | 0.342 | **-0.012** | 0.004 | **-0.010** | 0.004 |
| April | -0.452 | 0.317 | 0.024 | 0.331 | **-0.009** | 0.004 | **-0.008** | 0.004 |
| May | -0.330 | 0.307 | -0.015 | 0.321 | **-0.008** | 0.004 | -0.007 | 0.004 |
| June | -0.193 | 0.298 | 0.033 | 0.313 | **-0.011** | 0.003 | **-0.009** | 0.003 |
| July | 0.005 | 0.283 | 0.760 | 0.298 | **-0.008** | 0.003 | -0.006 | 0.003 |
| August | 0.125 | 0.275 | 0.778 | 0.288 | -0.005 | 0.003 | -0.004 | 0.003 |
| September | -0.214 | 0.266 | 0.335 | 0.278 | -0.002 | 0.003 | -0.001 | 0.003 |
| October | -0.132 | 0.242 | 0.495 | 0.253 | -0.002 | 0.003 | -0.001 | 0.003 |
| November | -0.160 | 0.216 | **0.560** | 0.228 | 0.000 | 0.003 | 0.001 | 0.003 |
| December | -0.100 | 0.182 | **0.401** | 0.192 | -0.001 | 0.002 | -0.001 | 0.002 |
| January | -0.138 | 0.129 | **0.429** | 0.138 | 0.001 | 0.002 | 0.001 | 0.002 |
| February | | | | | | | | |
| March (passage) | 0.138 | 0.136 | **1.076** | 0.145 | -0.003 | 0.002 | **-0.003** | 0.002 |
| April | **0.791** | 0.209 | **0.960** | 0.217 | -0.005 | 0.002 | **-0.005** | 0.002 |
| May | **0.818** | 0.258 | **1.192** | 0.267 | -0.004 | 0.002 | -0.004 | 0.002 |
| June | **1.156** | 0.287 | **1.500** | 0.296 | **-0.006** | 0.003 | **-0.007** | 0.003 |
| July | **1.408** | 0.304 | **1.743** | 0.315 | **-0.016** | 0.003 | **-0.016** | 0.003 |
| August | **1.840** | 0.326 | **2.612** | 0.337 | **-0.024** | 0.004 | **-0.024** | 0.004 |
| September | **2.047** | 0.349 | **2.240** | 0.359 | **-0.034** | 0.004 | **-0.033** | 0.004 |
| October (full impl) | **2.605** | 0.363 | **2.896** | 0.373 | **-0.045** | 0.004 | **-0.044** | 0.004 |
| November | **3.415** | 0.375 | **3.815** | 0.386 | **-0.053** | 0.004 | **-0.053** | 0.004 |
| December | **3.390** | 0.383 | **3.345** | 0.394 | **-0.064** | 0.004 | **-0.062** | 0.004 |
| January | **3.279** | 0.383 | **3.342** | 0.394 | **-0.074** | 0.004 | **-0.072** | 0.004 |
| February | **3.465** | 0.399 | **3.979** | 0.411 | **-0.080** | 0.005 | **-0.077** | 0.005 |
| March | **3.911** | 0.405 | **3.546** | 0.416 | **-0.082** | 0.005 | **-0.079** | 0.005 |

*Notes:* This table shows estimates corresponding to those in Figures 3,4 and 6.

Table A7

Estimated effects of the reform by month for mothers; any hours, part time (at least 80 hours per month), or full time work (at least 160 hours per month)

| | Outcome: Any hours worked<br>Comparative event study | | Outcome: At least 80 hours worked<br>Comparative event study | | Outcome: At least 160 hours worked<br>Comparative event study | |
|---|---|---|---|---|---|---|
| | Coefficient | Standard error | Coefficient | Standard error | Coefficient | Standard error |
| January | 0.000 | 0.003 | -0.001 | 0.002 | -0.001 | 0.001 |
| February | -0.001 | 0.003 | -0.001 | 0.002 | -0.002 | 0.001 |
| March | -0.002 | 0.003 | -0.003 | 0.002 | -0.001 | 0.001 |
| April | -0.004 | 0.003 | -0.003 | 0.002 | *-0.002* | 0.001 |
| May | -0.003 | 0.003 | -0.003 | 0.002 | 0.000 | 0.001 |
| June | -0.001 | 0.003 | -0.003 | 0.002 | 0.000 | 0.001 |
| July | -0.002 | 0.003 | -0.001 | 0.002 | 0.001 | 0.001 |
| August | -0.001 | 0.003 | 0.000 | 0.002 | 0.002 | 0.001 |
| September | -0.003 | 0.002 | -0.002 | 0.002 | -0.002 | 0.001 |
| October | -0.002 | 0.002 | -0.001 | 0.002 | 0.000 | 0.001 |
| November | -0.001 | 0.002 | -0.001 | 0.002 | -0.001 | 0.001 |
| December | -0.001 | 0.002 | -0.001 | 0.001 | 0.000 | 0.001 |
| January | -0.002 | 0.001 | -0.001 | 0.001 | 0.000 | 0.001 |
| February | 0.002 | 0.001 | 0.001 | 0.001 | 0.000 | 0.001 |
| March (passage) | | | | | | |
| April | **0.004** | 0.002 | **0.005** | 0.002 | *0.001* | 0.001 |
| May | **0.009** | 0.002 | **0.005** | 0.002 | -0.001 | 0.001 |
| June | **0.015** | 0.002 | **0.007** | 0.002 | *-0.002* | 0.001 |
| July | **0.021** | 0.003 | **0.009** | 0.002 | -0.002 | 0.001 |
| August | **0.028** | 0.003 | **0.010** | 0.002 | -0.002 | 0.001 |
| September | **0.031** | 0.003 | **0.011** | 0.002 | -0.002 | 0.002 |
| October (full impl) | **0.039** | 0.003 | **0.014** | 0.003 | -0.001 | 0.002 |
| November | **0.045** | 0.003 | **0.019** | 0.003 | 0.001 | 0.002 |
| December | **0.045** | 0.003 | **0.020** | 0.003 | 0.001 | 0.002 |
| January | **0.046** | 0.003 | **0.019** | 0.003 | 0.001 | 0.002 |
| February | **0.049** | 0.003 | **0.020** | 0.003 | 0.001 | 0.002 |
| March | **0.050** | 0.004 | **0.023** | 0.003 | **0.005** | 0.002 |

*Notes:* This table shows the estimates from a comparative event study estimation anchored in February, just prior to the passage of the reform. 2016 (2015) population consists of 18,578 (19,716) mothers on welfare benefits in March 2016 (2015).

Table A8

Ten most common jobs held

| | Share |
|---|---|
| Aide in elderly care | 0.169 |
| Cleaning assistant, offices or residential homes | 0.116 |
| Peadogical assistant | 0.092 |
| Sales agent | 0.066 |
| Clerk | 0.042 |
| Kitchen assistant | 0.036 |
| Cash register operator | 0.036 |
| Nursing aide, hospitals or institutions | 0.026 |
| Warehouse assistant | 0.018 |
| Teacher, primary or lower secondary schools | 0.013 |

*Notes:* This table shows the ten most common jobs held in 2016 for the group of mothers exposed to the reform.

Table A9

Estimated effects of the reform: Channels and potential mechanisms

| | Coef. | Std. error |
|---|---|---|
| Family adjusted discretionary income (€) | | |
| One year after | **-316.06** | 5.57 |
| | | |
| # mothers in comparison group | 18,578 | |
| # mothers in treatment group | 19,716 | |
| | | |
| Change of address | | |
| One year after | **0.011** | 0.004 |
| | | |
| # mothers in comparison group | 18,578 | |
| # mothers in treatment group | 19,716 | |
| | | |
| Absence rate | | |
| 1st quarter after (Q2) | **0.012** | 0.001 |
| 2nd quarter after (Q3) | 0.001 | 0.001 |
| 3rd quarter after (Q4) | 0.002 | 0.001 |
| 4th quarter after (Q1) | **0.007** | 0.002 |
| | | |
| # children in comparison group | 27,529 | |
| # children in treatment group | 25,666 | |
| | | |
| Any accidents | | |
| 1st quarter after (Q2) | -0.001 | 0.002 |
| 2nd quarter after (Q3) | **0.006** | 0.002 |
| 3rd quarter after (Q4) | 0.001 | 0.002 |
| 4th quarter after (Q1) | -0.002 | 0.002 |
| | | |
| # children in comparison group | 36,859 | |
| # children in treatment group | 39,415 | |

*Notes:* The table shows the results from comparative event study estimation. The reform (comparison) cohort consists of families (family adjusted discretionary income), mothers (change of address), children enrolled in public shools (absence rate), and all children of mothers on welfare (accidents) in March 2016 (2015). The analysis is anchored just prior to the reform, i.e., in March 2016 (2015 for the comparison cohort) for family adjusted discretionary income and change of address and in Q1 for the absence rate. **Bold** indicates significance at a 5% level; *italic* indicates significance at a 10% level.

Table A10

Estimated effects of the reform on each subquestion on the social well-being survey

| | Coefficient estimate | Standard error | Sample size |
|---|---|---|---|
| How well do you like your school? | **-0.075** | 0.020 | 45,782 |
| How well do you like the other children in your classroom? | **-0.067** | 0.020 | 45,703 |
| Do you feel lonely? (reverse coded) | **0.053** | 0.020 | 45,338 |
| Are you afraid of being ridiculed at school? (reverse coded) | -0.017 | 0.020 | 45,886 |
| Do you feel safe at school? | **-0.070** | 0.020 | 45,372 |
| Since the start of the school year, did anyone bully you? (reverse coded) | **-0.068** | 0.021 | 44,726 |
| I feel I belong at my school. | **-0.113** | 0.020 | 44,725 |
| I like the breaks at school. | **-0.118** | 0.025 | 45,208 |
| Most of the pupils in my classroom are kind and helpful. | -0.009 | 0.023 | 44,949 |
| Other pupils accept me as I am. | **-0.077** | 0.022 | 45,278 |

*Notes:* The table shows the results from comparative event study estimation using 2015-2017 data for reform cohort and 2015-2016 data for comparison cohort. The reform (comparison) cohort consists of children of mothers on welfare in March 2016 (2015) who were enrolled in public schools. The analysis is anchored in Q1, just prior to the passage of the reform. Post-measurement is Q1 2017 (2016) for the reform (comparison) cohort. Model controls for linear time trend but this does not impact the estimated effect of the reform. **Bold** indicates significance at a 5% level; *italic* indicates significance at a 10% level.

Table A11

Reports to child protective services:

Reasons for concerns and types of informants, by sex of the child

| | Share: | | |
|---|---|---|---|
| | Overall | Boys | Girls |
| Reason for concern \| concern | | | |
| Drug abuse, child | 0.016 | 0.018 | 0.014 |
| Crime, child | 0.051 | 0.083 | 0.023 |
| Problems at school, e.g., absence | 0.078 | 0.077 | 0.080 |
| Other child problem behaviors; e.g, externalizing behaviors | 0.216 | 0.219 | 0.213 |
| Disability, child | 0.027 | 0.029 | 0.025 |
| Abuse (sexual, violence) towards child | 0.071 | 0.067 | 0.077 |
| Other type of abuse or neglect | 0.092 | 0.091 | 0.092 |
| Drug abuse, parents | 0.095 | 0.085 | 0.108 |
| Crime, parents | 0.010 | 0.009 | 0.011 |
| Disability, parents | 0.072 | 0.066 | 0.079 |
| High level of conflict or violence between adults at home | 0.126 | 0.123 | 0.129 |
| Insufficient care from parents | 0.157 | 0.152 | 0.163 |
| Residential tenant eviction, homelessness | 0.051 | 0.046 | 0.057 |
| Other reason | 0.225 | 0.222 | 0.229 |
| | | | |
| Type of informant \| reason for concern: | | | |
| School | 0.206 | 0.218 | 0.193 |
| Health care provider | 0.129 | 0.115 | 0.147 |
| Anonymous | 0.081 | 0.076 | 0.088 |
| Relative, child in question, or acquaintance | 0.076 | 0.078 | 0.073 |
| Municipal transfer in connection with moves | 0.081 | 0.080 | 0.083 |
| Police or court | 0.080 | 0.093 | 0.065 |
| Day care institution | 0.068 | 0.075 | 0.061 |
| Other | 0.266 | 0.254 | 0.279 |
| | | | |
| # reports | 15,299 | 8,272 | 7,119 |

*Notes:* This table shows reasons and types of informants for report to child protective services for 2016 for the group of children exposed to the reform.

Table A12

Estimated effects of the reform by quarter, reports to child protective services

| | Coefficient | Standard error |
|---|---|---|
| 1st quarter after | **0.008** | 0.003 |
| 2nd quarter after | 0.004 | 0.003 |
| 3rd quarter after | **0.010** | 0.003 |
| 4th quarter after | **0.019** | 0.003 |

*Notes:* This table shows estimates corresponding to those in Figure 5.

Table A13

Ten most common injuries

| ICD-10 code | | Share |
|---|---|---|
| S60 | Superficial injury of wrist and hand (like contusion of a finger) | 0.091 |
| S01 | Open wound of head | 0.090 |
| S93 | Dislocation, sprain and strain of joints and ligaments at ankle and foot level | 0.078 |
| S52 | Fracture of forearm | 0.059 |
| S63 | Dislocation, sprain and strain of joints and ligaments at wrist and hand level | 0.059 |
| S90 | Superficial injury of ankle and foot (like contusion of an ankle) | 0.053 |
| S62 | Fracture at wrist and hand level | 0.051 |
| S00 | Superficial injury of head | 0.046 |
| S50 | Superficial injury of forearm (like contusion of elbow) | 0.041 |
| S61 | Open wound of wrist and hand | 0.039 |

*Notes:* This table shows the ten most common injuries that led to hospital visits in 2016 among children of the group of mothers exposed to the reform.